\documentclass[12pt,final]{article}
\usepackage{hyperref}
\usepackage{url,color}
\usepackage[final]{graphicx}  
\usepackage[dvipsnames]{xcolor} 
\usepackage{subfig}           
\usepackage{microtype}        
\usepackage[noabbrev]{cleveref} 
\usepackage{amsmath}
\usepackage{dsfont}
\usepackage[utf8]{inputenc} 
\usepackage{commath}          
\microtypesetup{
  activate={true,nocompatibility},  
  kerning=true,
  spacing=true,
  protrusion=true
}
\microtypecontext{spacing=nonfrench}
\makeatletter
\newif\if@preliminary
\@preliminaryfalse
\def\preliminary{\@preliminarytrue}

\def\preprintno#1{\def\@preprintno{#1}}
\def\address#1{\def\@address{#1}}
\def\email#1#2{\thanks{\tt #1@{}#2}}
\def\abstract#1{\def\@abstract{#1}}
\renewcommand\abstractname{ABSTRACT}
\newlength\preprintnoskip
\newlength\abstractwidth
\@titlepagetrue
\renewcommand\maketitle{\begin{titlepage}%
  \let\footnotesize\small
  \hfill\parbox{\preprintnoskip}{%
  \begin{flushright}\@preprintno\end{flushright}}\hspace*{1cm}
  \vskip 60\p@
  \begin{center}%
    {\Large\bf\boldmath \@title \par}\vskip 1cm%
    {\sc\@author \par}\vskip 3mm%
    {\@address \par}%
    \if@preliminary
      \vskip 2cm {\large\sf PRELIMINARY DRAFT \par \@date}%
    \fi
  \end{center}\par
  \@thanks
  \vfill
  \begin{center}%
    \parbox{\abstractwidth}{\centerline{\abstractname}%
    \vskip 3mm%
    \@abstract}
  \end{center}
  \end{titlepage}%
  \setcounter{footnote}{0}%
  \let\thanks\relax\let\maketitle\relax
  \gdef\@thanks{}\gdef\@author{}\gdef\@address{}%
  \gdef\@title{}\gdef\@abstract{}\gdef\@preprintno{}
}%
\def\@citex[#1]#2{\if@filesw\immediate\write\@auxout{\string\citation{#2}}\fi
  \def\@citea{}\@cite{\@for\@citeb:=#2\do
    {\@citea\def\@citea{,\penalty\@m}\@ifundefined
       {b@\@citeb}{{\bf ?}\@warning
       {Citation `\@citeb' on page \thepage \space undefined}}%
\hbox{\csname b@\@citeb\endcsname}}}{#1}}
\def\citerange{\@ifnextchar [{\@tempswatrue\@citexr}{\@tempswafalse\@citexr[]}}
\def\@citexr[#1]#2{\if@filesw\immediate\write\@auxout{\string\citation{#2}}\fi
  \def\@citea{}\@cite{\@for\@citeb:=#2\do
    {\@citea\def\@citea{--\penalty\@m}\@ifundefined
       {b@\@citeb}{{\bf ?}\@warning
       {Citation `\@citeb' on page \thepage \space undefined}}%
\hbox{\csname b@\@citeb\endcsname}}}{#1}}
\long\def\@makecaption#1#2{%
  \vskip\abovecaptionskip
  \sbox\@tempboxa{#1: \emph{#2}}%
  \ifdim \wd\@tempboxa >\hsize
    #1: \emph{#2}\par
  \else
    \hbox to\hsize{\hfil\box\@tempboxa\hfil}%
  \fi
  \vskip\belowcaptionskip}
\makeatother

\newcommand{\ii}{\mathrm{i}}

\newcommand{\MM}{\mathcal{M}}
\newcommand{\Aj}{\mathcal{A}^j}
\newcommand{\TT}{\mathcal{T}}

\newcommand{\vT}{\mathbf{T}}

\newcommand{\vA}{\mathbf{A}}

\newcommand{\TeV}{\text{TeV}}

\makeatletter
\font\manfnt=manfnt
\def\Watchout{\@ifnextchar [{\W@tchout}{\W@tchout[1]}}
\def\W@tchout[#1]{{\manfnt\@tempcnta#1\relax%
  \@whilenum\@tempcnta>\z@\do{%
    \char"7F\hskip 0.3em\advance\@tempcnta\m@ne}}}
\def\remark{\@ifnextchar[{\@remark}{\@remark[1]}}

\def\@remark[#1]#2{%
  \setbox\@tempboxa\hbox{\W@tchout[#1]}
  \@tempdima\wd\@tempboxa
  \list{}%
    {\leftmargin\@tempdima}%
    \item[\hbox to 0pt{\hss\W@tchout[#1]}]%
    \textbf{[#2]}}
\makeatother

\begin{document}


\preprintno{%
  KA-TP-15-2018\\
}

\title{
Anomalous quartic gauge couplings and unitarization for the vector boson scattering process $pp\rightarrow W^+W^+jjX\rightarrow \ell^+\nu_\ell\ell^+\nu_\ell jjX$
  }

\author{
  Genessis Perez\email{genessis.perez}{kit.edu}$^a$,
  Marco Sekulla\email{marco.sekulla}{kit.edu}$^a$,
  Dieter Zeppenfeld\email{dieter.zeppenfeld}{kit.edu}$^a$
}

\address{\it%
    $^a$Institute for Theoretical Physics, Karlsruhe Institute of
    Technology, D--76128 Karlsruhe, Germany
}

\date{%
  \today
}

\abstract{%
Weak vector boson scattering (VBS) at the LHC provides an excellent source of 
information on the structure of quartic gauge couplings and possible 
effects of physics beyond the SM in electroweak symmetry breaking. 
Parameterizing deviations from the SM 
within an effective field theory at tree level, the dimension-8 operators, 
which are needed for sufficiently general modeling, lead to unphysical 
enhancements of cross sections within the accessible energy range of the LHC.
Preservation of unitarity limits is needed for phenomenological studies of 
the $VVjj$ events which signify VBS. 
Here we develop a numerical unitarization scheme for the full off-shell VBS
processes and apply it to same-sign $W$ scattering, i.e. processes like
$qq\to qqW^+W^+$. The scheme is implemented within the Monte Carlo program 
VBFNLO, including leptonic decay of the weak bosons and NLO 
QCD corrections. Distributions differentiating between higher dimensional 
operators are discussed.  }
\maketitle




\section{Introduction}
\label{sec:intro}

Among the scattering processes which can be studied at the CERN Large Hadron
Collider (LHC),
weak vector boson scattering (VBS) is particularly interesting as a probe of 
electroweak symmetry breaking. Within the Standard Model (SM), intricate 
cancellations between Feynman amplitudes involving quartic gauge boson 
interactions, trilinear gauge boson couplings, and Higgs exchange lead to 
scattering amplitudes for longitudinally polarized weak bosons which do 
not grow with energy and which, for a light Higgs boson, respect bounds 
derived from unitarity. Modifications of the
weak boson couplings, among themselves or to the Higgs boson, spoil these 
cancellations and can lead to sizable cross section increases. For example,
reduced weak boson couplings to the light, $m_h=125$~GeV Higgs boson and 
compensation by an additional heavy Higgs in a two-Higgs-doublet model 
would lead to a cross section increase at high energy, as would a change 
only in the quartic gauge couplings.

  \begin{figure}[th!]
    \begin{center}
      \includegraphics[width=0.6\linewidth]{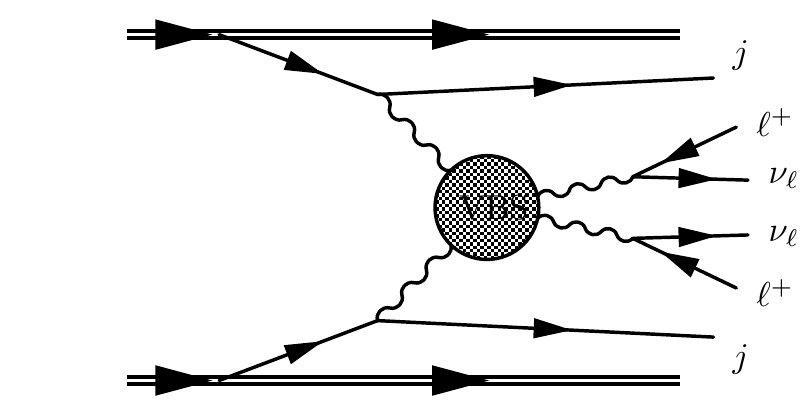} 
      \caption{Vector boson scattering contribution to the process 
               $pp \rightarrow W^+W^+ jjX$.}
      \label{fig:process}
    \end{center}
  \end{figure}  
Absent clear hints for a particular theory beyond the Standard Model (BSM),
a bottom up approach is conveniently formulated within an effective field 
theory (EFT) approach~\cite{Georgi:1994qn,Degrande:2012wf}. 
  Given the observation of a light Higgs boson at the 
  LHC~\cite{Aad:2012tfa,Chatrchyan:2012xdj}, we opt for a linear 
  representation of the light fields 
  in order to construct dimension-six and -eight operators for the EFT. 
  To give just one example, a deviation in the Higgs sector could manifest 
  itself via the dimension-8 term
  \begin{align}
    \label{eq:FS_1}
    \mathcal{L}_{S_1}  = \frac{f_{S_1}}{\Lambda^4}
      \left[ (D_\mu \phi )^\dagger D^\mu \phi \right] \left [(D_\nu \phi )^\dagger D^\nu \phi\right ] \, ,
  \end{align}
  in the effective Lagrangian, which is present in the linear 
  \'Eboli-basis~\cite{Eboli:2006wa}. Here, the covariant derivative 
  of the Higgs-doublet field, $D^\mu \phi$, contains $W$ and $Z$ fields, 
  $\Lambda$, is the energy scale of new physics, and the coupling 
  coefficient $f_{S_1}$ is used later to allow different strengths for 
  independent dimension-8 operators. 
  This operator will induce an anomalous contribution to the four-$W$, 
  four-$Z$, and $WWZZ$ vertices, which alter the scattering (predominantly)
  of the 
  longitudinal degrees of freedom of the weak vector bosons.  
The impact of anomalous couplings on $VV\to VV$ scattering can be 
studied at the LHC via the full process $pp \rightarrow V  V jjX$ as 
illustrated in Fig.~\ref{fig:process}, where the two final state vector bosons 
$V$ can decay either leptonically or hadronically. 

  The current, observed limits for ${f_{S_1}/\Lambda^4}$, derived from 
  same-sign $W$ scattering by CMS, are
  $\left[-21.6,21.8 \right] \; \TeV^{-4}$ for 35.9~fb$^{-1}$ of 
  $\sqrt{s}=13 \; \TeV$ data~\cite{Sirunyan:2017ret}. We are not aware of
  new results for Run-II published by ATLAS. However, comparing old limits for 
  ${f_{S_1}/\Lambda^{-4}}$ of $\left[-118,120 \right] \; \TeV^{-4}$ from 
  19.4~fb$^{-1}$ of $\sqrt{s}=8 \; \TeV$ data from 
  CMS~\cite{Khachatryan:2014sta} and $\left[-960,960 \right] \; \TeV^{-4}$ from 
  20.3~fb$^{-1}$ of $\sqrt{s}=8\;\TeV$ data from ATLAS~\cite{Aaboud:2016ffv}, 
  one observes a substantial difference in precision.\footnote{The limit 
  in Ref.~\cite{Aaboud:2016ffv} is determined with coefficients $\alpha_4,
  \alpha_5$ of the non-linear basis defined in \cite{Appelquist:1993ka}. 
  We used the conversion given in \cite{Sekulla:2016yku,Rauch:2016pai} 
  to transform these into limits of the linear \'Eboli basis.} 
  This difference is mainly due to the different high-energy extrapolation 
  of the EFT ansatz in the generation of BSM Monte-Carlo events. The EFT 
  is only valid up to a certain energy scale $\Lambda_{\mathrm{valid}}<\Lambda$, 
  where the operator product
  expansion breaks down. However, the experiment is only sensitive to the
  ratio ${f_{S_1}/\Lambda^4}$ and the scale $\Lambda$ is a priori not known.
  Using just the EFT as input for the generation of Monte Carlo data will    
  usually overshoot any result allowed by perturbative unitarity in the high 
  energy region. Naturally this will result in
  more stringent limits for the EFT-coefficients. CMS is using this approach in
  presenting their limits. ATLAS on the other hand is using the
  T-matrix~\cite{Kilian:2014zja,Kilian:2015opv} unitarization scheme to provide
  a theoretically consistent description of the high energy region, 
  where unitarity would otherwise be violated, with a proper 
  interpolation to the low energy EFT. T-matrix unitarization leads to 
  lower generated event rates for a given ${f_{S_1}/\Lambda^4}$, which leads to 
  weaker limits for this Wilson coefficient. 

  \begin{figure}[bt]
    \centering
    \includegraphics[height= 8.5 cm, page=1]{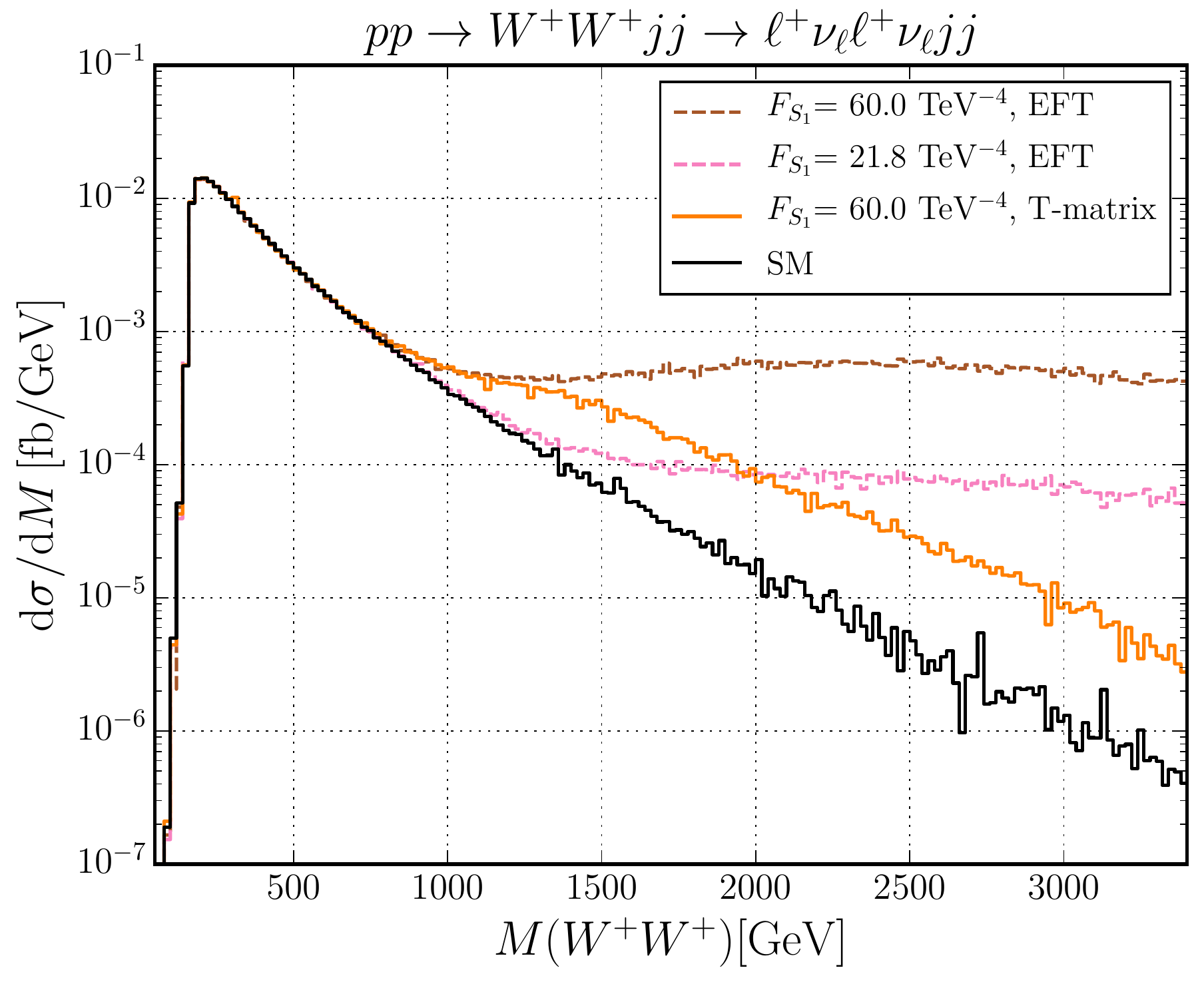}
    \caption{
      Differential cross section as a function of the invariant mass $m_{VV}$ of
      the weak vector bosons for $pp \rightarrow \ell^+ \nu_\ell \ell^+ \nu_\ell
      jjX$. The solid black line represents the SM, while the dashed pink and
      dashed brown lines show the EFT cross section for the 
      $F_{S1}={f_{S_1}/\Lambda^4}$ anomalous
      coupling. The solid orange line shows the T-matrix unitarized curve with
      the same fiducial cross section as the orange pure EFT curve. Cuts 
      defining the fiducial region are given in Eq.~\eqref{eq:cuts}.  }
     \label{fig:intro_mvv}
  \end{figure}

The effect is demonstrated in Fig.~\ref{fig:intro_mvv}, where we compare 
the $W^+W^+$ invariant mass distribution expected with the current CMS 
limit $F_{S_1}={f_{S_1}/\Lambda^4}=21.8$~TeV$^{-4}$ for a naive EFT 
description (dashed pink), with a T-matrix unitarized~\cite{Kilian:2014zja} 
prescription, with larger coupling $F_{S_1}=60$~TeV$^{-4}$, which will give 
the same fiducial cross section (solid orange histogram). Also shown 
are the SM expectation (solid black) and the naive EFT expectation 
for the larger coupling of 60~TeV$^{-4}$ (dashed brown), which agrees with 
the T-matrix unitarized expectation only at small invariant masses. In the 
energy range above $M_{WW}\approx 1500$~GeV, an $\mathcal{L}_{S_1}$ induced
excess above the orange $F_{S_1}=60$~TeV$^{-4}$ curve violates unitarity, 
i.e. it is unphysical, and should therefore 
not be considered to estimate EFT coefficients.
This is quite general: a pure effective 
Lagrangian/anomalous coupling analysis of LHC observables, with a finite set 
of terms in the effective Lagrangian, is insufficient in practice 
because the unbounded growth of amplitudes with energy typically corresponds 
to unitarity violation within the energy reach of the LHC. We thus need a 
general and versatile unitarization procedure for the naive EFT amplitudes 
at high momentum transfers, which smoothly interpolates to the pure EFT 
description well below $\Lambda_{\mathrm{valid}}$.

In order to analyze $VVjj$ production data, any unitarity considerations 
for $VV\to VV$ scattering must be extrapolated from on-shell bosons to the 
space-like incoming and time-like outgoing virtualities of the vector bosons 
which is implicit in the kinematics of 
Fig.~\ref{fig:process}. As we shall see, this extrapolation will require a 
few additional assumptions and will induce some model dependence.
To obtain predictions which are compatible with unitarity, the T-matrix 
unitarization prescription can be used.
So far, however, an implementation of this scheme is only available for a small
number of effective Lagrangian operators for VBS
due to the difficulty to handle VBS with arbitrarily polarized off-shell
vector bosons in the full $VVjj$ production 
process~\cite{Kilian:2014zja,Kilian:2015opv}.
In this paper, we introduce a variant of the T/K-matrix unitarization
scheme \cite{Kilian:2014zja,Alboteanu:2008my}, called $T_u$~unitarization below, for general combinations of
operators within 
VBS, for arbitrary space-like virtualities of the incoming vector bosons, 
and describe its implementation in the Monte Carlo generator 
VBFNLO~\cite{Arnold:2008rz,Arnold:2011wj,Baglio:2014uba}.

The paper is organized as follows. We introduce the full set of 
bosonic dimension-8 operators with a list of current experimental limits 
in Section~\ref{sec:EFT}. 
In Section~\ref{sec:unitarity}, we first consider how unitarity relations
can be extended to off-shell VBS processes. Beyond the definition of 
off-shell polarization vectors, this entails partial wave decomposition 
for off shell sub-amplitudes for $VV\to VV$ scattering and its 
fast numerical implementation. The $T_u$ 
unitarization model, which we have implemented in VBFNLO for same-sign $W$ 
scattering, is introduced in Section~\ref{sec:unitarisation}. 
Section~\ref{sec:results} is devoted to numerical results for same-sign 
$W$-scattering, i.e the process 
$pp \rightarrow W^+W^+ jj X\rightarrow \ell^+ \nu_\ell \ell^+ \nu_\ell jj X$ 
at NLO QCD precision, which is now implemented in VBFNLO including 
$T_u$ unitarization for any combination of the dimension-8 operators listed 
in Section~\ref{sec:EFT}. We compare our $T_u$ unitarized model with
naive EFT descriptions for different dimension-8 operators. Furthermore, we
will also give examples of observables helping to distinguish experimentally
between different subclasses of dimension-8 operators. 
Final conclusions are drawn in Section~\ref{sec:conclusions}.

\section{Effective Field Theory Description of \\ Anomalous Quartic Gauge Couplings}
\label{sec:EFT}
The bottom-up EFT framework is useful to quantify deviations from the SM in 
a model independent way and, once experimental evidence for such deviations 
is discovered, it gives hints, from which BSM 
effect a possible anomaly might originate. Short of such a desirable situation,
experimental limits on the Wilson coefficients serve as a measure of the
experimental precision. Two EFT representations are mainly used to describe
BSM contributions for anomalous quartic gauge couplings, the linear and
non-linear representation. They can be distinguished due to the different
ordering of the EFT expansion
\begin{align}
 \mathcal{L}_{\mathrm{EFT}} &=  \sum_i \frac{f_i}{\Lambda^{d_i-4}} \mathcal{O}_i\,,
\end{align}
which is written in terms of operators $\mathcal{O}_i$ of energy 
dimension $d_i$, corresponding 
Wilson coefficients $f_i$ (which allow for variations in importance of 
the individual operators) and energy scale of new physics, $\Lambda$. 
In the non-linear representation, the Higgs
couplings are treated as additional free parameters and deviations in the Higgs
sector can already be introduced at lowest order~\cite{Buchalla:2013rka}. This
was well motivated before the experimental Higgs discovery in case of a heavy or
strongly interacting Higgs~\cite{Appelquist:1980vg,Longhitano:1980iz}. However, 
since no deviations from the light SM Higgs predictions have been observed 
so far, we choose the linear Higgs representation, where deviations from 
the SM predictions for Higgs couplings and trilinear or quartic gauge boson 
first appear at energy 
dimension $d_i=6$~\cite{Buchmuller:1985jz,Hagiwara:1993ck,Grzadkowski:2010es}. 

Anomalous quartic gauge couplings (aQGC) are induced at the dimension-6 level 
already. However, they are not independent of changes in the Higgs couplings 
or of anomalous trilinear gauge couplings. These three-boson couplings are 
most easily measured in Higgs production or decay or in vector boson pair 
production ($q\bar q\to V_1V_2$), at the LHC, and little additional 
information is to be expected from the measurement of the significantly 
smaller VBS cross sections. Also, the tensor structure of 
dimension-6 operators is not general enough to allow for sufficiently 
uncorrelated variations of the 81 helicity amplitudes which, in principle, 
can be probed in a VBS process, $V_1V_2\to V_3V_4$, with massive vector bosons.
In this paper, we study aQGC which
enter the EFT at lowest order at dimension-8 without contributing 
to anomalous trilinear gauge interactions or to $HVV$ couplings. 

The
contributing CP conserving operators can be assembled from three SM building
blocks. One building block is the covariant derivative acting on the Higgs
doublet field,
\begin{equation}
  D_\mu \Phi \equiv 
\left( \partial_\mu + i \frac{g'}{2} B_\mu  + i g W_\mu^i \frac{\tau^i}{2} \right) \Phi,
  \label{eq:def_dmu}
\end{equation}
which affects the coupling of longitudinal modes of the gauge bosons.
Here, the Higgs, $H$ is embedded in the Higgs doublet field in the 
unitary gauge:
\begin{align}
  \Phi &= \begin{pmatrix}
    0 \\ \frac{v+ H}{\sqrt{2}} 
  \end{pmatrix} \, .
\end{align}
The other building blocks are the field strength tensors 
\begin{subequations}
 \label{eq:fields}
  \begin{align}
    \widehat{W}_{\mu\nu} & =  ig\frac{\tau^i}{2} 
    (\partial_\mu W^i_\nu - \partial_\nu W^i_\mu
    - g \epsilon_{ijk} W^j_\mu W^k_\nu ) \, ,\\ 
    \widehat{B}_{\mu \nu} & = \frac{i}{2} g' 
    (\partial_\mu B_\nu - \partial_\nu B_\mu)  \; ,
  \end{align}
\end{subequations}
which are normalized such that 
$[D_\mu,D_\nu]=\widehat{W}_{\mu\nu}+\widehat{B}_{\mu \nu}$ for the
covariant derivative in Eq.~\eqref{eq:def_dmu}. The abelian parts of these 
field strength tensors lead to 
couplings of the transverse degrees of freedom of the gauge fields.

The dimension-8 operators are separated into longitudinal, transverse, and mixed
contributions, corresponding to the occurrence of the building blocks above. A
revised list of dimension-8 operators from \cite{Eboli:2006wa} and 
\cite{Baak:2013fwa} is given in 
Eqs.~(\ref{eq:dim8-longitudinal},\ref{eq:dim8-mixed},\ref{eq:dim8-transversal}).
In comparison to 
the operators defined in \cite{Eboli:2006wa}, we choose a different 
normalization for the field strength in Eq.~\eqref{eq:fields}, which is 
accompanied by an additional factor of $\ii g$ or $\ii g^\prime/2$. 
These normalization choices are labeled
as "\'Eboli" for \cite{Eboli:2006wa} and the normalization in 
Eq.~\eqref{eq:fields} as "VBFNLO", in the following.

For the longitudinal operators the two normalization choices coincide:
\begin{subequations}
  \label{eq:dim8-longitudinal}
  \begin{alignat}{2}
    {\cal O}_{S_0} &= \left [ \left ( D_\mu \Phi \right)^\dagger
      D_\nu \Phi \right ] &&\times
    \left [ \left ( D^\mu \Phi \right)^\dagger
      D^\nu \Phi \right ] \, ,\\
    {\cal O}_{S_1} &= \left [ \left ( D_\mu \Phi \right)^\dagger
      D^\mu \Phi  \right ] &&\times
    \left [ \left ( D_\nu \Phi \right)^\dagger
      D^\nu \Phi \right ] \, \\
      {\cal O}_{S_2} &= \left [ \left ( D_\mu \Phi \right)^\dagger
      D_\nu \Phi \right ] &&\times
    \left [ \left ( D^\nu \Phi \right)^\dagger 
      D^\mu \Phi \right ] \, .
  \end{alignat}
\end{subequations}
Compared to Ref.~\cite{Eboli:2006wa}, the longitudinal operator set is 
extended by the operator $\mathcal{O}_{S_2}$, which is needed for a simultaneous 
matching to the non-linear basis for all weak boson flavor combinations 
in VBS~\cite{Rauch:2016pai,Baak:2013fwa}. The mixed set is given by 
\begin{subequations}
  \label{eq:dim8-mixed}
\allowdisplaybreaks
\begin{align}
  {\cal O}_{M_0} &=   \hbox{Tr}\left [ {\widehat{W}}_{\mu\nu} {\widehat{W}}^{\mu\nu} \right ]
  \times  \left [ \left ( D_\beta \Phi \right)^\dagger
    D^\beta \Phi \right ]  
  \, ,\\
  {\cal O}_{M_1} &=   \hbox{Tr}\left [ {\widehat{W}}_{\mu\nu} {\widehat{W}}^{\nu\beta} \right ]
  \times  \left [ \left ( D_\beta \Phi \right)^\dagger
    D^\mu \Phi \right ]  
  \, ,\\
  {\cal O}_{M_2} &=   \left [ {\widehat{B}}_{\mu\nu} {\widehat{B}}^{\mu\nu} \right ]
  \times  \left [ \left ( D_\beta \Phi \right)^\dagger
D^\beta \Phi \right ]  
\, ,\\
{\cal O}_{M_3} &=   \left [ {\widehat{B}}_{\mu\nu} {\widehat{B}}^{\nu\beta} \right ]
\times  \left [ \left ( D_\beta \Phi \right)^\dagger
  D^\mu \Phi \right ]  
\, ,\\
{\cal O}_{M_4} &= \left [ \left ( D_\mu \Phi \right)^\dagger {\widehat{W}}_{\beta\nu}
  D^\mu \Phi  \right ] \times {\widehat{B}}^{\beta\nu}  
\, ,\\
{\cal O}_{M_5} &= \left [ \left ( D_\mu \Phi \right)^\dagger {\widehat{W}}_{\beta\nu}
  D^\nu \Phi  \right ] \times {\widehat{B}}^{\beta\mu}  
\, , \\
{\cal O}_{M_5^\prime} &= \left [ \left ( D_\mu \Phi \right)^\dagger {\widehat{W}}^{\beta\mu}
  D^\nu \Phi  \right ] \times {\widehat{B}}_{\beta\nu}  
\, , \\
{\cal O}_{M_7} &= \left [ \left ( D_\mu \Phi \right)^\dagger {\widehat{W}}_{\beta\nu}
  {\widehat{W}}^{\beta\mu} D^\nu \Phi  \right ]  \, .
\end{align}
\end{subequations}
The operator ${\mathcal {O}}_{M_6}$ of the original operator set 
in~\cite{Eboli:2006wa} is not independent of the others 
(${\mathcal {O}}_{M_0}=2{\mathcal {O}}_{M_6}$) and can therefore 
be omitted. We have added $\mathcal {O}_{M_5^\prime}$,
which is the hermitian conjugate of $\mathcal {O}_{M_5}$, and has to be included
to complete the operator set. Finally, the purely transverse operators are
\begin{subequations}
  \label{eq:dim8-transversal}
  \begin{alignat}{2}
    {\cal O}_{T_0} &=   \hbox{Tr}\left [ {\widehat{W}}_{\mu\nu} {\widehat{W}}^{\mu\nu} \right ]
    &&\times   \hbox{Tr}\left [ {\widehat{W}}_{\alpha\beta} {\widehat{W}}^{\alpha\beta} \right ]  
    \, ,\\
    {\cal O}_{T_1} &=   \hbox{Tr}\left [ {\widehat{W}}_{\alpha\nu} {\widehat{W}}^{\mu\beta} \right ] 
    &&\times   \hbox{Tr}\left [ {\widehat{W}}_{\mu\beta} {\widehat{W}}^{\alpha\nu} \right ]  
    \, ,\\
    {\cal O}_{T_2} &=   \hbox{Tr}\left [ {\widehat{W}}_{\alpha\mu} {\widehat{W}}^{\mu\beta} \right ]
    &&\times   \hbox{Tr}\left [ {\widehat{W}}_{\beta\nu} {\widehat{W}}^{\nu\alpha} \right ]   
    \, ,\\
    {\cal O}_{T_5} &=   \hbox{Tr}\left [ {\widehat{W}}_{\mu\nu} {\widehat{W}}^{\mu\nu} \right ]
    &&\times   {\widehat{B}}_{\alpha\beta} {\widehat{B}}^{\alpha\beta}   
    \, ,\\
    {\cal O}_{T_6} &=   \hbox{Tr}\left [ {\widehat{W}}_{\alpha\nu} {\widehat{W}}^{\mu\beta} \right ]
    &&\times   {\widehat{B}}_{\mu\beta} {\widehat{B}}^{\alpha\nu}   
    \, ,\\
    {\cal O}_{T_7} &=   \hbox{Tr}\left [ {\widehat{W}}_{\alpha\mu} {\widehat{W}}^{\mu\beta} \right ]
    &&\times   {\widehat{B}}_{\beta\nu} {\widehat{B}}^{\nu\alpha}   
    \, ,\\
    {\cal O}_{T_8} &=   {\widehat{B}}_{\mu\nu} {\widehat{B}}^{\mu\nu}  {\widehat{B}}_{\alpha\beta} {\widehat{B}}^{\alpha\beta} &&
    \, ,\\
    {\cal O}_{T_9} &=  {\widehat{B}}_{\alpha\mu} {\widehat{B}}^{\mu\beta}   {\widehat{B}}_{\beta\nu} {\widehat{B}}^{\nu\alpha} && \, .
  \end{alignat}
\end{subequations}

Same-sign $W$ boson scattering is the VBS process which can be measured 
with the highest precision, due to a sizable signal cross section and a 
particularly low QCD 
background~\cite{Sirunyan:2017ret,Khachatryan:2014sta,Aaboud:2016ffv}. 
In Section~\ref{sec:results} we will concentrate on this process, to which
only operators with exactly four $W^\pm$ fields can contribute at tree level. 
This eliminates
all operators with a hypercharge field strength, ${\widehat{B}}_{\mu\nu}$. The 
remaining ones have been probed by ATLAS~\cite{Aaboud:2016ffv} and 
CMS~\cite{Sirunyan:2017ret,Khachatryan:2014sta} in same-sign $W$ scattering, 
and the results are summarized in Table~\ref{table:limits}.

\begin{table}[h]
\begin{center}
\begin{tabular}{c| c c c c}
 Measurement &CMS, 13 TeV\cite{Sirunyan:2017ret}  
 &CMS, 13 TeV &ATLAS, 8 TeV\cite{Aaboud:2016ffv}  
 &CMS, 8 TeV\cite{Khachatryan:2014sta}     \\
 Normalization & \'{E}boli & 
 VBFNLO & VBFNLO (T-matrix) & \'{E}boli\\ \hline
$f_{S_0}/\Lambda^4$ &[-7.7,7.7]  &[-7.7,7.7]        &                           &[-38,40]      \\
$f_{S_1}/\Lambda^4$ &[-21.6,21.8]&[-21.6,21.8]      &[-960,960]        &[-118,120]    \\
$f_{M_0}/\Lambda^4$ &[-6.0,5.9]  &[-14,15]&                           &[-33,32]      \\
$f_{M_1}/\Lambda^4$ &[-8.7,9.1]  &[-22,21]&                           &[-44,47]      \\
$f_{M_6}/\Lambda^4$ &[-11.9,11.8]&[-28.7,28.9]&                           &[-65,63]      \\
$f_{M_7}/\Lambda^4$ &[-13.3,12.9]&[-31.4,32.3]&                           &[-70,66]      \\
$f_{T_0}/\Lambda^4$ &[-0.62,0.65]&[-3.7,3.8]  &                           &[-4.2,4.6]    \\
$f_{T_1}/\Lambda^4$ &[-0.28,0.31]&[-1.7,1.8]  &                           &[1.9,2.2]     \\
$f_{T_2}/\Lambda^4$ &[-0.89,1.02]&[-5.3,6.0]  &                           &[-5.2,6.4]    \\
\end{tabular}

\end{center}
\caption{Experimental limits (in TeV$^{-4}$) on the coefficients of 
dimension-8 operators, $f_i/\Lambda^4$, from observation of 
$pp~\rightarrow~W^\pm W^\pm~jjX$.}
\label{table:limits} 
\end{table}
In comparing the different 
normalization of \'Eboli and VBFNLO, one finds that the limits for VBFNLO 
differ by about one order of magnitude only, whereas the limits in the 
\'Eboli normalization vary by up to two orders of magnitude. The difference 
is simply due to consistently factorizing the small electroweak couplings, 
which are expected for any model explaining the EFT, into the definition 
of the operators for the VBFNLO normalization convention.
For the 8~TeV data, the ATLAS result incorporates a unitarization model 
to prevent the generation of unphysical events at high energy, which violate
unitarity constraints. The corresponding bound on 
$f_{S_1}/\Lambda^4$ for the same-sign W scattering
process observed by ATLAS \cite{Aaboud:2016ffv} is approximately one order of
magnitude weaker than the CMS $\sqrt{s}=8 \,\mathrm{TeV}$ limit, which 
indicates the impact that unitarization can have on quoted 
experimental results.

\section{Unitarity for VBS: going off-shell }
\label{sec:unitarity}

\begin{figure}[ht!] 
  \begin{center}
  \subfloat[Vector boson emission\label{subfig:feynman-emission}]{%
  \includegraphics[width=0.45\linewidth]{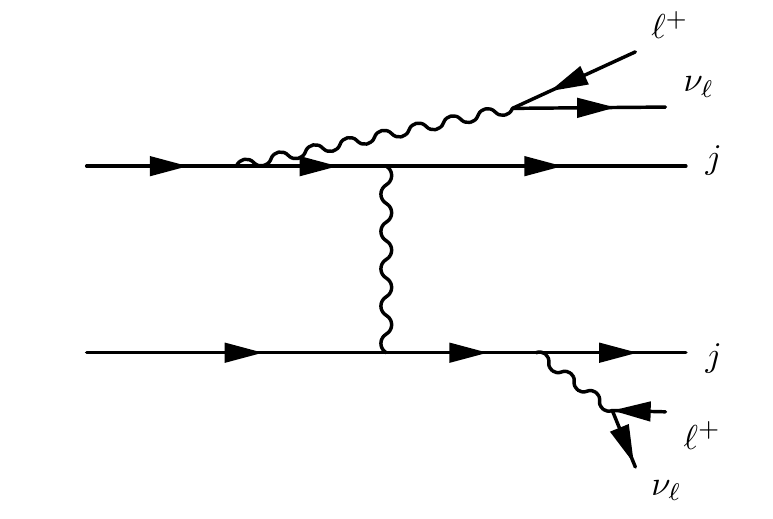} 
  }
  \subfloat[Quartic gauge interaction.\label{subfig:feynman-quartic}]{%
  \includegraphics[width=0.45\linewidth]{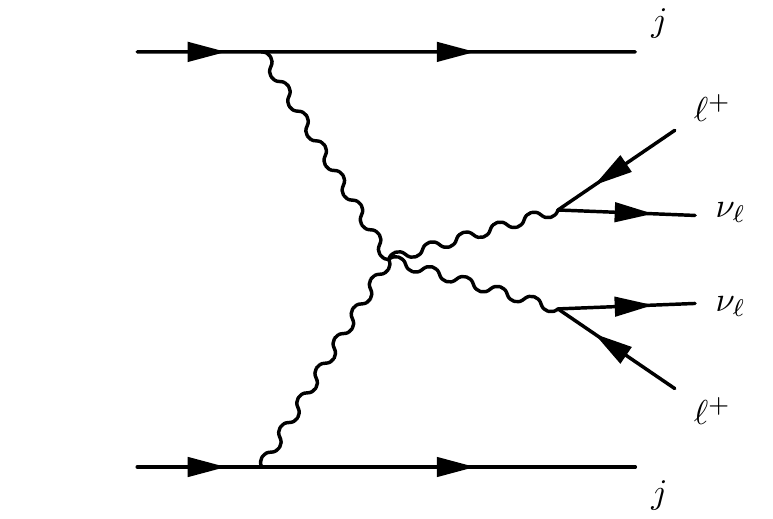} 
  }  
\end{center}
\caption{Examples of Feynman graphs contributing to vector boson scattering.}
\label{fig:feynman}
\end{figure}

We need to apply unitarity considerations to electroweak processes of the 
type $pp \rightarrow \bar{\psi_1}\psi_2\bar{\psi_3}\psi_4 jj$
at ${\cal O}(\alpha^6)$ (LO) and at ${\cal O}(\alpha^6\, \alpha_s)$ (NLO), i.e.
including QCD corrections. At the parton level, the $\psi_i$ represent  
decay leptons of two vector bosons, the initial $pp$ state represents the 
scattering partons (quarks or anti-quarks in the LO case) and $jj$ stands 
for the final state partons yielding two tagging jets. Representative 
Feynman graphs for the 8-fermion processes at LO are 
given in Fig.~\ref{fig:feynman} and include vector boson emissions off 
quark lines
as in Fig.~\ref{subfig:feynman-emission} as well as VBS contributions as in  
Fig.~\ref{subfig:feynman-quartic}.
The BSM physics, which we consider via the introduction of bosonic operators, 
will only contribute to the VBS subprocess $VV\rightarrow VV$. The SM 
contributions to the complete process are gauge invariant by themselves, they
are ``small'' and they respect perturbative unitarity.
Splitting the full amplitude into the SM and a BSM piece,
\begin{align} 
\label{eq:amplitude_overall}
\MM_{pp\rightarrow 4f jj} &= \MM_{pp\rightarrow 4f jj}^{\rm SM} + 
\MM_{pp\rightarrow 4f jj}^{\rm BSM} \, ,
\end{align}
it is, therefore, sufficient to unitarize the BSM piece only, via the VBS 
subprocess, which means that we neglect the interference of SM and 
BSM amplitudes for unitarization.\footnote{As we will see, unitarized 
cross sections exceed SM expectation 
by more than an order of magnitude, which justifies this approximation.}

\subsection{Identification of the $VV\rightarrow VV$ Subamplitude}
\label{sec:subAmplitude}
Within the VBFNLO approach, the entire $VV\rightarrow VV$ subprocess is 
contained inside a leptonic tensor, which then is contracted with quark 
currents, $J^\mu_{p \rightarrow j V}$. These represent the emission of a virtual 
vector boson, $V$, off an initial parton in 
Fig.~\ref{subfig:feynman-quartic}. This 
structure is well suited to implement the BSM amplitude as 
\begin{align}
\label{eq:amplitude_BSM}
\MM_{pp\rightarrow 4f jj}^{\rm BSM} &= 
J^\mu_{p_1 \rightarrow j V_1} J^\nu_{p_2 \rightarrow j V_2} \, 
\MM_{V_1V_2\rightarrow 4f\, \mu\nu }^{\rm BSM} \, ,
\end{align}
where $\MM_{V_1V_2\rightarrow 4f\, \mu\nu }^{\rm BSM}$ denotes the BSM 
contribution to the leptonic tensor.

The quark currents, $J^\mu_{p \rightarrow j V}$, and the decay currents,
$J_{V \rightarrow \bar{f}f}^\mu$, are conserved, since we are neglecting 
fermion masses, and this allows for a simple expansion of the off-shell 
vector boson propagators in terms of polarization vectors of fixed helicity.
When writing  
\begin{align} \label{eq:amplitude_heldec}
\MM_{pp\rightarrow 4f jj}^{\rm BSM} = & 
J_{p_1 \rightarrow j V_1}^\mu \ J_{p_2 \rightarrow j V_2}^\nu \ 
D_{\mu\alpha}^{V_1}(q_1) \  D_{\nu\beta}^{V_2}(q_2) \nonumber \\
& \times \textnormal{\bf M}_{V_1 V_2 \rightarrow V_3 V_4}^{\alpha\beta\gamma\delta}  
\  D_{\gamma\rho}^{V_3}(q_3) \  D_{\delta\sigma}^{V_4}(q_4) \\
& \times J_{V_3 \rightarrow \bar{f}f}^\rho \  J_{V_4 \rightarrow \bar{f}f}^\sigma  \, ,
\nonumber
\end{align}   
the vector boson propagators may be taken as
\begin{align}
\label{eq:propagator_sum_rule}
D^{\mu\nu}_{V}(q) = \frac{-\ii}{q^2 - m_V^2+i\,m_V\,\Gamma_V}\left(g^{\mu\nu} - \frac{q^\mu q^\nu}{q^2} \right) \equiv
\frac{-\ii}{q^2-m_V^2+i\,m_V\,\Gamma_V}\sum_\lambda\epsilon_J^{* \, \mu}(q,\lambda)
  {\epsilon}_{\MM}^{\nu}(q,\lambda) \, 
\end{align}
since the $q^\mu$ terms are contracted with a conserved current 
and, thus, vanish.
The indices $J$ and $\MM$ on the polarization vectors distinguish 
between those that are contracted with the currents, $\epsilon_J$,
and the polarization vectors contracted with the VBS matrix element,
$\epsilon_\MM$. Furthermore, we generalize the definition of
the polarization vectors for off-shell vector bosons with four-momentum $q$ to
\begin{subequations}
\label{eq:def-polarization-vectors}
\begin{align}
\epsilon^\mu_{J}(q,\pm) &=
   \mp\frac{1}{\sqrt{2}\sqrt{q_x^2+q_y^2}}
      \left(0; \frac{q_zq_x}{|\vec q|} \mp \ii q_y,
               \frac{q_yq_z}{|\vec q|} \pm \ii q_x,
               - \frac{q_x^2+q_y^2}{|\vec q|}\right)
            =\epsilon^\mu_{\MM}(q,\pm) \, ,\\
\epsilon^\mu_{J}(q,0) &=
  \mathcal{N}_J \left(|\vec{q}| , \, q_0 \frac{\vec{q}}{|\vec{q}|} \right) \,
 , \qquad \qquad \qquad\qquad
\mathcal{N}_\MM \left(|\vec{q}| , \, q_0 \frac{\vec{q}}{|\vec{q}|} \right)
 =\epsilon^\mu_{\MM}(q,0) \, ,
\end{align}
\end{subequations}
where $\mathcal{N}_\MM$ and $\mathcal{N}_J$ have to fulfill
$\mathcal{N}_J \mathcal{N}_\MM = 1/q^2$ as normalization factors
for the longitudinal polarization vectors. 
One could choose the individual factors to be equal in magnitude. However, 
in order to match the proper normalization for on-shell weak bosons and thus to 
reproduce the correct normalization of $VV\rightarrow VV$ scattering 
amplitudes for longitudinal $V$, we set   
\begin{align}
\mathcal{N}_J = \frac{m_V}{q^2} \, , \qquad \mathcal{N}_\MM = \frac{1}{m_V} .
\end{align}

With these definitions, the BSM contribution to the leptonic tensor becomes
\begin{align}
\label{eq:amplitude_decomposed}
\MM_{V_1V_2\rightarrow 4f, \mu\nu }^{\rm BSM} = &
\prod_{i=1}^4 \frac{1}{q_i^2-m_{V_i}^2+i\,m_{V_i}\,\Gamma_{V_i}} 
\sum_{\{\lambda_i\}}
  \epsilon^*_{J,\mu}(q_1,\lambda_1)\epsilon^*_{J,\nu}(q_2,\lambda_2)
  \MM^{VBS}_{\lambda_3,\lambda_4;\lambda_1,\lambda_2} \\ \nonumber
  & \epsilon_{J}(q_3,\lambda_3)\cdot J_{V_3 \rightarrow \bar{f_1}f_2} \quad
  \epsilon_{J}(q_4,\lambda_4)\cdot J_{V_4 \rightarrow \bar{f_3}f_4} \, .
\end{align}
The full anomalous VBS information is contained in the helicity amplitudes
\begin{align}
\MM^{VBS}_{\lambda_3,\lambda_4;\lambda_1,\lambda_2}\left(q_3,q_4;q_1,q_2 \right) \,
&= \epsilon_{\MM,\alpha}(q_1,\lambda_1) \epsilon_{\MM,\beta}(q_2,\lambda_2) \, 
  \textnormal{\bf M}_{V_1 V_2 \rightarrow V_3 V_4}^{\alpha\beta\gamma\delta} \, 
  \epsilon^*_{\MM,\gamma}(q_3,\lambda_3) \epsilon^*_{\MM,\delta}(q_4,\lambda_4) \, .
\end{align}
For on-shell vector boson momenta, they correspond to the normal 
$VV\to VV$ helicity amplitudes induced by the dimension-8 operators. 
For the full $pp \rightarrow 4f jj$ process, however, we are dealing with 
incoming space-like four-momenta $q_1$ and $q_2$, and 
outgoing time-like four-momenta $q_3$ and $q_4$, which means that the initial 
and final states of even the elastic $W^+W^+\to W^+W^+$ process do not 
properly match. 

In the presence of dimension-8 operators, the tree level VBS amplitudes
$\MM^{VBS}$ can rise with the fourth power of the center-of-mass energy, 
$\sqrt{s}$. For example, the BSM part of the helicity amplitude
$\MM^{VBS}_{+-;+-}$, for the operator $\mathcal{O}_{T_0}$, is given by
\begin{align}
  \label{eq:analyticalAmplitude}
  \MM^{VBS}_{+-;+-} &=
  \frac{f_{T_0}}{\Lambda^4}  2 g^4 \cos^4\left( \frac{\Theta}{4} \right)
  \left[s^2 - s(q_1^2+q_2^2 -2m_W^2) + 4 m_W^2(q_1^2+q_2^2) - \frac{2 m_W^2}{s}
  (q_1^2-q_2^2)^2 \right] ,
\end{align}
where the time-like momenta are approximated as on-shell, $q_3^2=q_4^2 \approx
m_W^2$. This example also shows that unphysical, strong enhancements are 
possible for large $s$ and for large $q_i^2$, independently.

In order to avoid unphysical behaviour, within the energy range probed
by the LHC, the subprocess amplitudes $\MM_{pp\rightarrow 4f jj}^{\rm BSM}$ of
Eq.~(\ref{eq:amplitude_BSM})  need to be replaced by unitarized versions, for
Wilson coefficients of practical interest. Since we intend to describe BSM
interactions of the known SM bosons, the unitarization has to act at the level
of $VV\to VV$ scattering instead of the full $pp\rightarrow 4f jj$ subprocess:
working at the latter level, unitarity alone would e.g. allow replacement of the
well-known narrow Breit Wigner propagator for the $W$ by a broad spectral
function, keeping the leptons produced in $W\to \ell\nu$ on top of resonance for
virtuality ranges of hundreds of GeV, which would also result in large cross
section increases. Our choice of the physics which we want to describe, 
forces us to match the offshell VBS amplitudes $\MM^{VBS}$ to unitarized 
on-shell  $VV\to VV$ scattering amplitudes, which can be defined from 
first principles. The guiding principles here are
\begin{itemize}
\item In the on-shell limit, $q_i^2\to m_{V_i}^2$, the unitarized off-shell 
amplitude must reduce to the corresponding unitarized on-shell amplitude.
\item For large virtualities, $q_i^2$, and modest $s$ (which is allowed for the 
incoming space-like bosons) the unitarized off-shell amplitude should not 
exceed the corresponding on-shell unitarity bound.
\item The unitarization procedure must reduce to the EFT limit when the 
absolute values of all invariants ($q_i^2$, and the Mandelstam variables, 
$s$, $t$ and $u$) are small compared to $\Lambda^2$, which sets the new 
physics scale.
\end{itemize}
These principles must now be applied to the unitarization of the off-shell 
VBS amplitudes.

\subsection{Unitarity Relation for $2\to 2$ Amplitudes}
\label{sec:unitarity-sub}
It is useful to briefly recall the derivation of unitarity relations as 
exposed, for example, in Ref.\cite{Itzykson:1980rh}. Starting point of any 
unitarization procedure is the unitarity of the 
scattering matrix $\mathbf{S}$
\begin{align}
  \mathbf{S}&= 1+ \ii \mathbf{T} \, ,\\
  2 \mathrm{Im} \mathbf{T}&= -\ii \left( \mathbf{T} - \mathbf{T}^\dagger\right)
  = \mathbf{T}^\dagger \mathbf{T} = \mathbf{T} \mathbf{T}^\dagger  \, ,
  \label{eq:unitarity_condition_T}
\end{align}
where, exploiting momentum conservation,  the $i\to f$ matrix elements
are given by
\begin{equation}
 \vT_{fi} =  (2\pi)^4 \delta(P_f-P_i) \, \TT_{fi}
\end{equation} 
Truncating the sum over intermediate states to the two-boson subspace, 
the elements of the $2\to2$ scattering matrix,
$\TT_{fi} (q_3,q_4 \leftarrow q_1, q_2)$, have to fulfill the condition
\begin{align}
  \TT_{fi}-\TT_{if}^* &= \ii \sum_n
    \int \underbrace{\frac{d^3{\bf q}_{n,3}d^3{\bf q}_{n,4}}
    {(2\pi)^3 2q^0_{n,3} (2\pi)^3 2q^0_{n,4}}
    (2\pi)^4 \delta(P_i-q_{n,3}-q_{n,4})}_
    {\frac{\lambda^{1/2}(s,q^2_{n,3},q^2_{n,4})}{8s(2\pi)^2}d\Omega}\, S_n \,
    \TT^*_{nf}\TT_{n i} \\
    2\mathrm{Im}(\TT_{fi}) &=
    \sum_n \frac{\lambda^{1/2}(s,q^2_{n,3},q^2_{n,4})}{8\pi s} \, S_n \,\int
    \frac{d\Omega}{4\pi}   \TT^*_{nf}\TT_{n i} \, , \label{eq:unitarity2b}
\end{align}
where $S_n$ is the statistical factor for identical particles, $S_n=\frac{1}{2}$
for the $W^+W^+$ case to be concentrated on later.
Exploiting angular momentum conservation, every $2\to 2$ helicity amplitude 
$\mathcal{M}_{\lambda_3\lambda_4\leftarrow\lambda_1\lambda_2}=\TT_{fi}$
can be expanded in corresponding partial wave amplitudes
$\mathcal{A}^j_{ \lambda_{3}\lambda_{4} \leftarrow \lambda_{1}\lambda_{2}}$
\begin{align}\label{eq:def-part-wave-amp}
  \mathcal{M}_{\lambda_3\lambda_4 \leftarrow \lambda_1\lambda_2} \left(\Theta, \varphi \right)
  &=
  8 \pi \mathcal{N}_{fi}
  \sum_{j=\operatorname{max}\left( \abs{\lambda_{12}}, \abs{\lambda_{34}} \right)}
    ^{j_{\mathrm{max}}}
    (2j +1) \mathcal{A}^j_{ \lambda_{3}\lambda_{4} \leftarrow \lambda_{1}\lambda_{2}}
    d^j_{ \lambda_{12}  \lambda_{34}} \left(\Theta \right)e^{\ii \lambda_{34}\varphi} \, ,
\end{align}
where $d^j_{ \lambda_{12}  \lambda_{34}}$ denotes a Wigner $d$-function,
$\lambda_{ij}= \lambda_i - \lambda_j$, and 
$\mathcal{N}_{fi}=\mathcal{N}_{fi}(q_3,q_4;q_1,q_2)$ is a 
normalization factor. Note that for the dimension-8 operators described 
in Section~\ref{sec:EFT}, only partial waves up to $j_{\mathrm{max}}=2$
contribute.\footnote{Since at most three partial waves contribute, 
  knowledge of the helicity amplitude $\mathcal{M}_{\lambda_3\lambda_4 \leftarrow
    \lambda_1\lambda_2}^{VBS}\left(\Theta\right)$ at three angles is
  sufficient to determine all partial wave amplitude
  $\mathcal{A}^j_{ \lambda_{3}\lambda_{4} \leftarrow \lambda_{1}\lambda_{2}}$ for a
  given set of helicities. We have implemented this procedure in VBFNLO.  }

Performing the angular integral in Eq.~(\ref{eq:unitarity2b}), 
the partial wave amplitudes $\mathcal{A}^j$ are found to satisfy the relation
\begin{align}\label{eq:partial-wave-unit}
  2\mathrm{Im}(\mathcal{A}^j_{ \lambda_{3}\lambda_{4} \leftarrow \lambda_{1}\lambda_{2}})
  &=
   \sum_n \frac{\mathcal{N}_{ni}\mathcal{N}_{nf}}{\mathcal{N}_{fi}} 
   \frac{\lambda^{1/2}(s,q_{n,3}^2,q_{n,4}^2)}{s}  \, S_n \,
  \sum_{\lambda_1^\prime,\lambda_2^\prime}
  {\mathcal{A}^j}^*_{ \lambda_{1}^\prime\lambda_{2}^\prime \leftarrow \lambda_{3}\lambda_{4} }
  \mathcal{A}^j_{\lambda_{1}^\prime\lambda_{2}^\prime \leftarrow \lambda_{1}\lambda_{2}}
\end{align}
where we have separated the sum over intermediate states into an explicit 
helicity sum and a sum over $n$, which corresponds to a sum over 
possible boson flavor combinations. Choosing
\begin{equation}\label{eq:part-wave-norm}
\mathcal{N}_{ni}= 
\frac{s}{\lambda^{1/4}(s,q_{n,3}^2,q_{n,4}^2)\lambda^{1/4}(s,q_{1}^2,q_{2}^2)} 
\frac{1}{\sqrt{S_nS_i}}
\end{equation} 
with K\"all\'en function 
\begin{align}
  \label{eq:Kaellen}
  \lambda (x_1,x_2,x_3)= x_1^2 +x_2^2+x_3^2 -2x_1x_2 -2x_1x_3 -2x_2x_3 \, ,
\end{align}
and analogously for $\mathcal{N}_{nf}$ and $\mathcal{N}_{fi}$, the phase-space 
factor in Eq.~(\ref{eq:partial-wave-unit}) is canceled, resulting in
a form analogous to Eq.~\eqref{eq:unitarity_condition_T}. Note 
that for the case at hand, $W^\pm W^\pm \to W^\pm W^\pm$ scattering,
the statistical factors are all equal,  $S_i=S_f=S_n=1/2$. However, the above 
description readily generalizes to more complex cases like 
$W^+ W^- \to W^+ W^-, \, ZZ, \, Z\gamma, \, HH$ etc..

Diagonalizing the partial wave helicity amplitudes, the eigenvalues $a^j(s)$ 
will lie on an Argand circle of radius unity, which implies 
\begin{equation}\label{eq:unit-bound}
|\mathrm{Re}(a^j(s))|\leq 1\, . 
\end{equation}
We will refer to this limit as the unitarity bound on the scattering amplitude. 
Alternatively one could use $|a^j(s)|\leq 2$, which is reached for a purely 
imaginary scattering amplitude. This comparison shows that the precise place 
at which a (real) tree level amplitude violates unitarity is somewhat ambiguous.
However, a polynomial growth with energy, as implied by a truncated EFT, 
is clearly forbidden by the unitarity relation of 
Eq.~(\ref{eq:partial-wave-unit}).

For on-shell $W^\pm W^\pm \to W^\pm W^\pm$ scattering the specification of 
virtualities in the above equations is superfluous, of course. However, 
we want to extend the formalism to the unitarization of the off-shell 
amplitudes $\MM^{VBS}$ of Eq.~(\ref{eq:amplitude_decomposed}), with 
space-like momenta $q_1$ and $q_2$ and time-like momenta $q_3$ and $q_4$
which are somewhat off the $W$ Breit-Wigner 
peak.\footnote{We have tried various options
of replacing the off-shell by on-shell helicity amplitudes, which form a 
normal scattering matrix~\cite{Perez2018}. However, one typically faces
significant cross section changes for large virtualities of the incoming 
vector bosons, which are not sufficiently suppressed by the propagators 
for dimension-8 operators or higher. Our solution avoids these problems.} 
For this general case,  
Eq.~(\ref{eq:def-part-wave-amp}) together with the normalization factor of 
Eq.~(\ref{eq:part-wave-norm}) defines the 
partial wave amplitudes to be used below.

Allowing free virtualities of the external particles leads to a new 
problem, however: already at tree level the scattering amplitudes 
$\mathcal{A}^j_{ \lambda_{3}\lambda_{4} \leftarrow \lambda_{1}\lambda_{2}}$ no longer form
normal matrices, i.e.  $\vT\vT^\dagger \neq \vT^\dagger\vT$ when states with 
different virtualities are identified, i.e. when they are associated with 
a single on-shell state. While the mismatch becomes sub-dominant for 
virtualities much smaller than the center of mass energy, i.e. 
for $\abs{q_i^2}/s \ll 1$, we here need an interpolation which 
also works for modest center of mass energies and virtualities, reproducing 
the EFT results, and which allows us to take the exact, off-shell 
helicity amplitudes $\MM^{VBS}$ as input for the unitarization
in all regions of phase space.
The proposed generalization will be described in the next section.

\subsection{Implementation of Unitarization: the $T_u$ Model}
\label{sec:unitarisation}

  Using a truncated EFT model at tree level for large energy scales will 
  violate unitarity above a certain energy. For current experimental limits 
  on the EFT coefficients, this unphysical behavior happens within the 
  energy reach of the LHC, as demonstrated in Fig.~\ref{fig:intro_mvv}. 
Therefore, an extended model must 
  be used to ensure that generated differential cross sections are not 
becoming unphysically large. Several 
  procedures, with different high energy behavior, are available to 
  extrapolate the EFT beyond its validity range. One possibility, 
which has been used in VBFNLO in the past, is the 
  introduction of (somewhat ad hoc) form-factors which multiply the full 
  BSM amplitude $\MM^{\rm VBS}$ of Eq.~(\ref{eq:amplitude_decomposed}) 
to ensure the unitary bound of Eq.~(\ref{eq:unit-bound}). 

Theoretically more attractive 
is the substitution of the tree level amplitudes by 
versions, which, at least approximately, satisfy the unitarity condition of
Eq.~(\ref{eq:partial-wave-unit}). One such procedure is the linear T-matrix 
projection for the intermediate $2 \to 2$ interaction matrix that is 
introduced in~\cite{Kilian:2014zja,Kilian:2015opv}. With this projection, 
the $2 \to 2$ scattering amplitudes will approach the perturbative unitarity 
bound at high energies and are matched to the naive EFT at low energies. 
Given the starting point of a normal\footnote{$\mathrm{Re}\mathbf{T}_0$,
  $\mathrm{Im}\mathbf{T}_0$, $\mathbf{T}_0$ and $\mathbf{T}_0^\dagger$ commute.}
tree level interaction matrix $\mathbf{T}_0$, the procedure corresponds to the 
substitution of $\mathbf{T}_0$ by  
  \begin{align}
    \mathbf{T}_L &= \left( \mathds{1} -\frac{\ii}{2} \mathbf{T}_0^{\dagger} \right)^{-1}
    \frac{1}{2} \left(\mathbf{T}_0 + \mathbf{T}_0^\dagger\right) \, .
    \label{eq:unit_linear}
  \end{align}

The T-matrix unitarization model has been implemented for the 
$\mathcal{O}_S$ operators~\cite{Kilian:2014zja,Kilian:2015opv,Loeschner2014}, 
which enhance the scattering of longitudinal vector 
bosons. In these implementations, an analytical approach has been chosen to
  provide T-matrix unitarized results at high center of mass energies. 
The next step in this program is the expansion of the
 method for operator classes $\mathcal{O}_M$ and $\mathcal{O}_T$, i.e. 
the implementation for additional helicity
  combinations of the vector bosons~\cite{Fleper:2018}. 

Contrary to the analytical ansatz chosen 
in~\cite{Kilian:2014zja,Kilian:2015opv,Loeschner2014}, which 
requires approximations which become exact only in the limit of 
$s\gg m_V^2, |q_i^2|$, we here opt for a numerical approach, which gives us 
greater versatility for the additional dimension-8 operator classes, 
allowing investigations of arbitrary regions of phase space. 
As mentioned 
in the introduction, we here limit ourselves to the doubly charged 
channels, i.e. to scattering of two same-sign $W^\pm$ bosons.

In the case of on-shell scattering, the interaction matrix becomes hermitian, 
at tree level, and we can expand the denominator in Eq.~\eqref{eq:unit_linear},
to improve numerical stability, as
  \begin{align}
    \mathbf{T}_L &= 
    \left( \mathds{1} +\frac{1}{4} \mathbf{T}_0\mathbf{T}_0\right)^{-1}
    \left( \mathbf{T}_0 +\frac{\ii}{2} \mathbf{T}_0\mathbf{T}_0  \right) \, .
    \label{eq:unit_linear_expanded}
  \end{align}
  As mentioned in the last section, the interaction matrix of the $2 \to 2$
  vector boson scattering subprocess, within the process $pp \rightarrow W^\pm
  W^\pm jj$, is not normal, because the momenta of the incoming vector 
bosons $q_1, q_2$
  are space-like and the momentum of the outgoing vector bosons $q_3,q_4$ are
  time-like and almost on-shell. Although an extended procedure for non-normal
  interaction matrices is provided in \cite{Kilian:2014zja}, it is not feasible
  for a numerical approach. 

To generalize Eq.~\eqref{eq:unit_linear_expanded} for off-shell sub-amplitudes  
$\MM^{\rm VBS}$,  we distinguish states with time-like and space-like bosons as 
separate classes, labeling the corresponding matrix elements with
$s$ for space-like and $t$ for time-like momenta. This leads us to consider
three cases for the 
partial wave 
amplitudes $\mathcal{A}^j$ defined in Eq.~(\ref{eq:def-part-wave-amp}),
\begin{subequations}  \label{eq:extended-interactionMatrix}
    \begin{align}
      \vA_{t\leftarrow s}&=\Aj_{\lambda_3,\lambda_4;\lambda_1,\lambda_2}(q_3,q_4;q_1,q_2) \, , \\
      \vA_{s\leftarrow t}&=\Aj_{\lambda_3,\lambda_4;\lambda_1,\lambda_2}(k_3,k_4;k_1,k_2) \, ,\\
      \vA_{t\leftarrow t}&=\Aj_{\lambda_3,\lambda_4;\lambda_1,\lambda_2}(q_3,q_4;k_1,k_2) 
\,, 
    \end{align}
\end{subequations}
which correspond to the amplitudes of the actual physical subprocess,
with time-like final momenta and space-like initial momenta, its hermitian 
adjoint, and an approximately on-shell amplitude, respectively. 
$\vA_{s\leftarrow s}$ is omitted, because a purely space-like 4-point function
does not appear as a sub-amplitude 
in a scattering process initiated by two particles only. 
The additionally introduced time-like momenta $k_1,k_2$ and space-like 
momenta $k_3,k_4$ in Eq.~\eqref{eq:extended-interactionMatrix} point in the 
same direction in 3-space as the original $q_i$, but with swapped virtualities.
More precisely, the invariant mass of the scattering weak boson pair, 
$\sqrt{s}$, is kept fixed and
  \begin{subequations}
    \begin{alignat}{2}
      & \vec{k}_i \parallel \vec{q}_i \, , &  \\ 
      k_1^2 = q_3^2 \, ,\quad & k_2^2 = q_4^2 \, , \quad  & k_3^2 = q_1^2 \, , \quad &
      k_4^2 = q_2^2 \, .
    \end{alignat}
  \end{subequations} 
We can identify the matrices of the right hand side of
Eq.~\eqref{eq:unit_linear_expanded} by following the guiding principles
introduced in section~\ref{sec:subAmplitude}. The matrix $\vT_0$ in the 
numerator has to be $\vA_{t\leftarrow s}$ to guarantee a reduction to the 
EFT limit for low energy scales. Additionally, the virtualities of 
polarization vectors in the sum over intermediate states have 
to be the same in order to guarantee reduction to the correct vector 
boson propagator (see Eq.~(\ref{eq:propagator_sum_rule})), i.e. in matrix 
multiplication of the helicity amplitudes in 
Eq.~(\ref{eq:extended-interactionMatrix}) only the products 
$\vA_{i\leftarrow t}\vA_{t\leftarrow j}$ or $\vA_{i\leftarrow s}\vA_{s\leftarrow j}$ 
are allowed. The unitarized interaction matrix has
to be of transition type $t \leftarrow s$ and, thus, the matrix product in the
  numerator is determined to be $\vA_{t\leftarrow t}\vA_{t\leftarrow s}$. The
  denominator has to behave as $t \leftarrow t$, which leaves only open the
  possibility of a linear combination of $\vA_{t\leftarrow t} \vA_{t\leftarrow
  t}$ and $\vA_{t\leftarrow s} \vA_{s\leftarrow t}$ for the matrix product in
  the denominator.

  This linear combination has to suppress both the polynomial rise with 
  the invariant mass of the scattering system, $\sqrt{s}$, as well as the
  rise with the space-like virtualities $q_1^2$ and $q_2^2$. Time-like 
virtualities are of no concern once the dependence on $s$ is addressed,
because $s$ provides an upper limit for $q_3^2$ and $q_4^2$. 
Contributions involving high virtuality space-like momenta, especially for the 
transverse operators, will eventually lead to a unphysical cross section growth
at $s\ll |q_1^2|\, , |q_2^2|$. 
An example is given in Eq.~\eqref{eq:analyticalAmplitude}. Either 
the $s\, q_{1/2}^2$ or the $q_{1/2}^4 m_W^2/s$ term could become dominant
at low $\sqrt{s}$. To ensure that the unitarized amplitude will not rise
due to un-suppressed space-like virtualities and therefore become unphysical, the
  denominator has to contain at least as many space-like states as 
  the numerator.
  Hence, the matrix product $\vA_{t\leftarrow t} \vA_{t\leftarrow t}$ has to be
  omitted and we arrive at the unitarization formula
  \begin{align}
    \label{eq:T-matrix-offshell}
      \vA^{\mathrm{unit}}_{t\leftarrow s} =&\left( \mathds{1} +
      \frac{1}{4}  \vA_{t\leftarrow s} \vA_{s\leftarrow t}
       \right)^{-1} 
       \left( \vA_{t\leftarrow s}+
       \frac{\ii}{2} \vA_{t\leftarrow t} \vA_{t\leftarrow s} 
        \right) \, .
  \end{align}

In this off-shell
extension of the linear T-matrix unitarization, the eigenvectors of 
denominator and numerator will only align exactly in the on-shell limit. 
In fact, since $\vA_{t\leftarrow s}$ is not normal, (non-aligned) eigenvectors 
can only be defined for the hermitian and the anti-hermitian parts of 
$\vA_{t\leftarrow s}$ separately. As a result, the suppression of large 
enhancements cannot be guaranteed for states which fall along eigenvectors of 
small eigenvalues of the denominator.
For the case at hand, $W^+W^+$ scattering, this is not problematic for the 
operators $\mathcal{O}_{S}$, where only 
one helicity combination, namely the purely longitudinal ones, will 
receive a leading contribution, proportional to $s^2$.
  However, multiple helicity combinations will receive a strong enhancement 
  if at least one coefficient of transverse or mixed dimension-8 operators 
is non-zero.
  Therefore, the formula in Eq. \eqref{eq:T-matrix-offshell} is still
not satisfactory. Using the maximal eigenvalue $a^2_{\mathrm{max}}$ of the matrix
product $\vA_{t\leftarrow s}  \vA_{s\leftarrow t}$ instead, individually 
for each $j=0,1,2$ partial wave, will ensure that the resulting
amplitudes are always below the unitarity limit. Our final 
unitarization formula, which we call the $T_u$ model, reads
  \begin{align}
    \label{eq:Tu-unitarisation}
\mathbf{T}_u =
      \vA^{\mathrm{unit}}_{t\leftarrow s} =&\left( \mathds{1} +
      \frac{1}{4} a^2_{\mathrm{max}}
       \right)^{-1} 
       \left( \vA_{t\leftarrow s}+
       \frac{\ii}{2} \vA_{t\leftarrow t} \vA_{t\leftarrow s} 
        \right) \, ,
  \end{align}
and fulfills all the guiding principles listed at the end of 
Section~\ref{sec:subAmplitude}. Note that other choices would be 
possible for the suppression factors $1/(1+a^2_{\mathrm{max}}/4)$. For example, 
$a^2_{\mathrm{max}}$ could be taken the same for the $j=0,1,2$ partial waves.
This would correspond to a common overall form-factor, i.e. the dynamical 
suppression of the EFT growth would set in at a unique scale of new physics
for all helicity combinations and partial waves. 
Clearly, such changes correspond to different models of the BSM
dynamics. Here, we use the $T_u$ model because it is closer to the previous
T-matrix unitarization model of Ref.~\cite{Kilian:2014zja}.

\section{Consequences for LHC Physics}
\label{sec:results}
Both the newly introduced $T_u$-model and T-matrix unitarization modify 
the naive EFT description in slightly different ways, but we expect both 
to agree at asymptotically large energies, $s\gg m_W^2,\, |q_i^2|$, when 
a single helicity configuration and, thus, a single large eigenvalue of 
the scattering matrix dominates the high energy behavior. In order 
to demonstrate these features, we 
start with a comparison of the three models, using Wilson coefficients 
near the present experimental limits for the dimension-8 operators. 
  Next we discuss their impact on various
  observables, with an eye to distinctions between the different operator
  classes. 
 
The Monte-Carlo generator VBFNLO is used to calculate distributions 
and fiducial cross sections for the vector boson scattering process 
$pp \rightarrow W^+W^+ jj X\rightarrow \ell^+ \nu_\ell \ell^+ \nu_\ell jj X$ 
at NLO QCD for $\sqrt{s}= 13 \, \mathrm{TeV}$. Here, $\ell^+$ denotes a 
positron or muon in the final state. The jets are defined by anti-$k_t$
clustering~\cite{Cacciari:2008gp} with radius $R=0.4$. They are ordered by
transverse momenta and the tagging jets at NLO are defined as the two hardest
jets. As default, we use the CT10 PDF set~\cite{Lai:2010vv}, and electroweak
parameters are determined within the $G_F$-scheme with the measured values of
$G_F$, $m_W$, $m_Z$ and $m_H$ as input. For the fiducial cross section we 
follow the recent CMS analysis~\cite{Sirunyan:2017ret} and use the 
following cuts, dubbed VBF cuts:
  \begin{align}
    \label{eq:cuts}
  \begin{alignedat}{3}
    m_{\ell \ell} &> 20 \,\mathrm{GeV}, \quad & m_{jj} &> 500 \,\mathrm{GeV}, \\
    p_T^\ell &> 20 \,\mathrm{GeV}, & p_T^j &> 30 \,\mathrm{GeV},& \quad p_T^{\mathrm{miss}} &> 30 \,\mathrm{GeV} \\
    |\eta_\ell| &< 2.5, & |\eta_j| &< 5,& \Delta\eta_{jj} &> 2.5\,.
  \end{alignedat}
  \end{align}

  \begin{figure}[!bt] 
    \subfloat[semi logarithmic scale\label{subfig:log}]{%
    \includegraphics[height= 6.55 cm, page = 1]{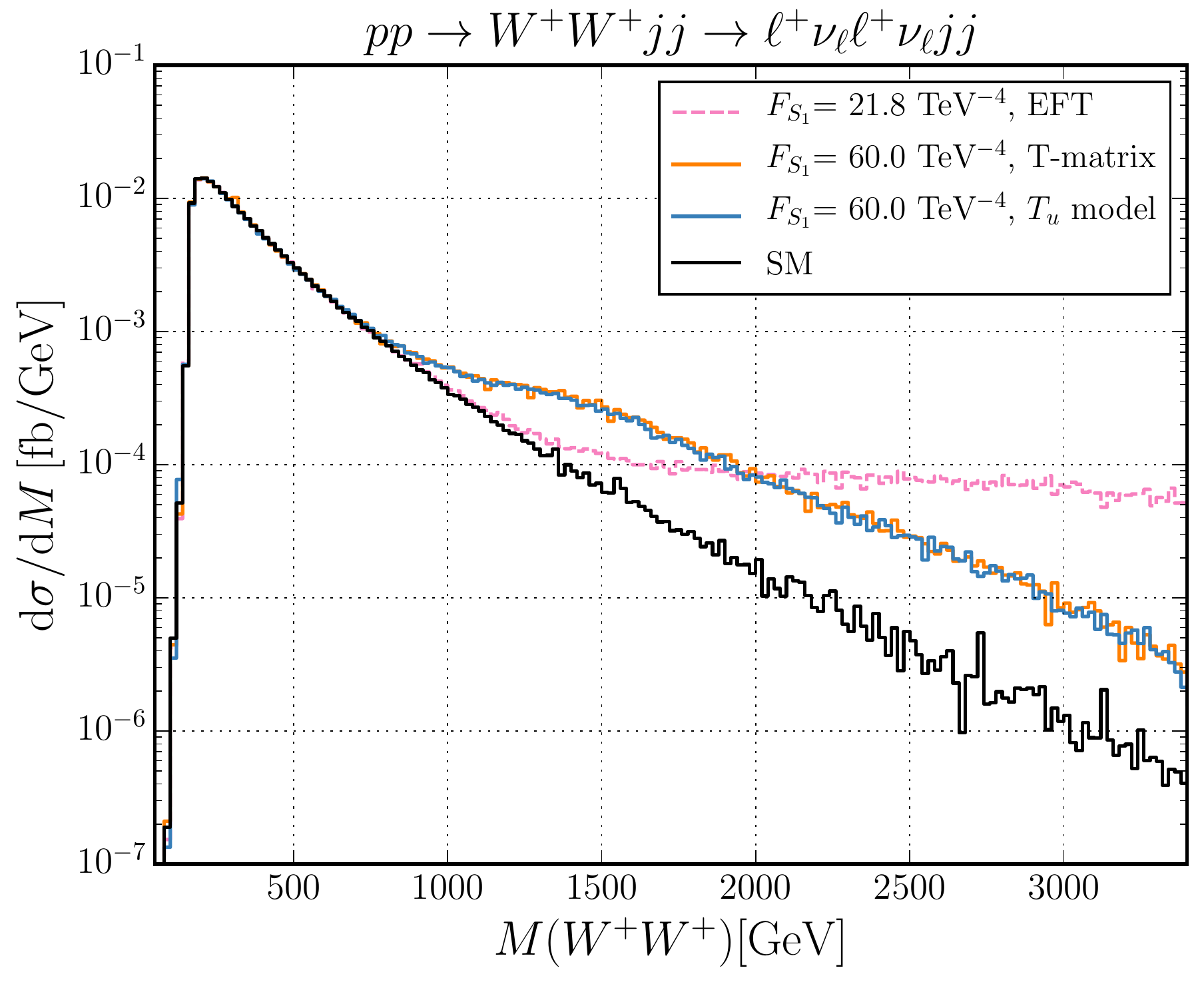}
    } 
    \hfill
    \subfloat[linear scale\label{subfig:linear}]{%
      \includegraphics[height= 6.55 cm, page = 1]{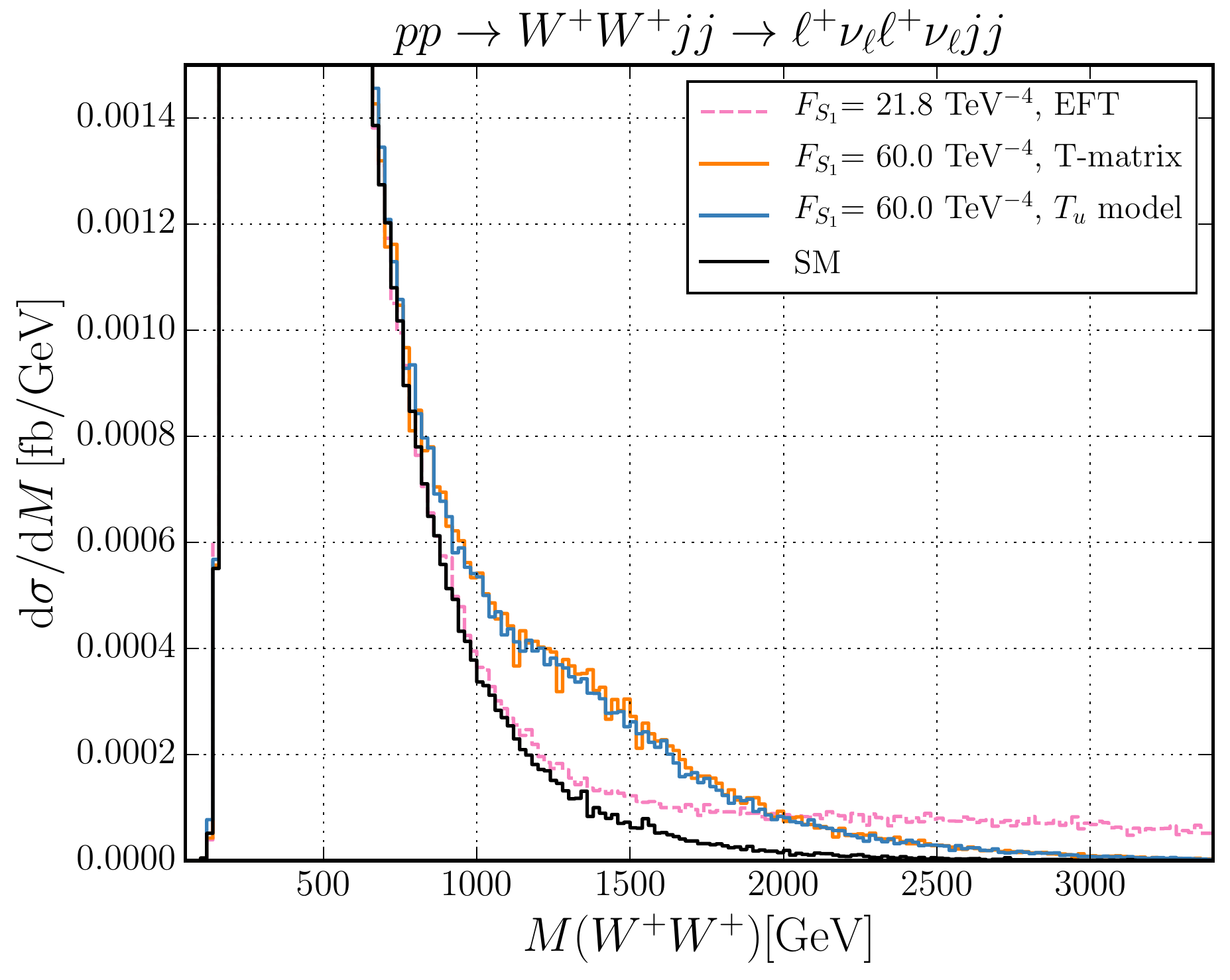}
    }
    \caption{ Comparison of $T_u$ model and linear T-matrix unitarization.
    Differential cross section as a function of the invariant mass $m_{WW}$ of
    the weak vector bosons for $pp\rightarrow\ell^+\nu_\ell\ell^+ \nu_\ell jjX$. 
    The solid black line shows the SM, while the dashed pink line shows the
    cross section for the $F_{S1}$ anomalous coupling using the present CMS 
    upper limit. The solid orange and blue lines show the a unitarized curve 
    with
    the same fiducial cross section as the EFT curve using the T-matrix and the
    $T_u$ model, respectively. Cuts defining the fiducial region are given 
    in Eq.~\eqref{eq:cuts}. }
     \label{fig:compare_UnitModel}
  \end{figure}
  
  In Fig.~\ref{fig:compare_UnitModel}, we compare the prediction of the naive
  EFT (dashed pink), the linear T-matrix unitarization (solid orange), the
  newly introduced $T_u$ model (solid blue) and the SM (solid black) for the
  longitudinal operator $\mathcal{O}_{S_1}$ as a function of the invariant mass
  of the vector boson pair. The coefficient for the naive EFT is chosen
  as the current experimental limit, 
  $F_{S_1}=f_{S_1}/\Lambda^4 = 21.8\, {\mathrm TeV}^{-4}$, of 
  CMS \cite{Sirunyan:2017ret} at $\sqrt{s}= 13 \, \mathrm{TeV}$. For the 
  unitarized models, T-matrix and $T_u$, we choose the coupling 
  such that the
  fiducial cross section of the naive EFT and the $T_u$ model coincide 
  within the VBF cuts of Eq.~(\ref{eq:cuts}). The number
  of produced events in the unitarized model are therefore nearly identical 
  to the naive EFT expectation within the high energy region used by CMS to 
  set the experimental bound, and thus the chosen value of 
  $F_{S_1}=60\, {\mathrm TeV}^{-4}$ approximates the present 
  bound on this coupling for the two unitarized models.
  Note that the coupling for the unitarized models is approximately a 
  factor 3 larger than for the naive EFT description. The expected excess of
  events with invariant masses above 2 TeV in the naive EFT description will
  violate unitarity and is therefore unphysical. Fig.~\ref{subfig:linear} shows
  that the events of this unphysical high energy tail need to be redistributed 
  to energies between 800 GeV and 2 TeV for the unitarized models, leading to 
  the weaker limit on the Wilson coefficient. Limits on 
  the dimension-8 coefficient derived with the naive EFT model overestimate 
  the sensitivity of experiments to the scale of high energy BSM effects.
  As displayed in Fig.~\ref{subfig:log}, the $T_u$ model reproduces the linear
  T-matrix unitarization prescription very well in the high energy 
  range, with barely visible differences at intermediate energies, below
  $M_{WW}\sim 2$~TeV, which can be traced to subleading effects in 
  $q_i^2/s$. The deviation of the unitarization models from the SM at high 
  energies is greatly reduced as compared to the naive EFT,
  and is a valid description beyond 
  the point of unitarity violation in the naive EFT model. 
  
  In Table~\ref{table:newlimits}, we list the estimated bounds on the full set 
  of dimension-8 coefficients for the $T_u$ model, derived as above by matching
  the fiducial cross section to the one obtained in the naive EFT model. We 
  stress that these numbers should be taken as rough estimates only, to be 
  superseded by full experimental analyses. Their main purpose here is for
  use in subsequent figures. They illustrate deviations from the SM which are
  at the edge of what is presently allowed experimentally.
  The bounds on these Wilson coefficients in the $T_u$ model are about a 
  factor of 3 weaker than the corresponding bounds derived within the naive 
  EFT, for 
  all three types of dimension-8 operators. Note, also, that the normalization 
  conventions differ between \'Eboli and VBFNLO definitions for mixed and 
  transverse operators. In the following we use the notation 
  $F_i = f^{\rm Eboli}_i/\Lambda^4$ for Wilson coefficients of operators 
  ${\cal O}_i$ defined with the \'Eboli normalization.  

  \begin{table}[h]
    \begin{center}
    \begin{tabular}{c| c c c c}
     Measurement &CMS, 13 TeV 
     &Corresponding $T_u$ &CMS, 13 TeV  
     &Corresponding $T_u$     \\
     Normalization & \'{E}boli & \'{E}boli &
     VBFNLO & VBFNLO \\ \hline
    $f_{S_0}/\Lambda^4$ &[-7.7,7.7]   &[-22,22]     &[-7.7,7.7]  & [-22,22] \\
    $f_{S_1}/\Lambda^4$ &[-21.6,21.8] & [-50,60]      &[-21.6,21.8] & [-50,60]\\
    $f_{M_0}/\Lambda^4$ &[-6.0,5.9]    & [-20.0,14.5] &[-14,15]    & [-35,49]  \\
    $f_{M_1}/\Lambda^4$ &[-8.7,9.1]   & [-29,23]  &[-22,21]    & [-56,71] \\   
    $f_{M_6}/\Lambda^4$ &[-11.9,11.8] & [-39,30]  &[-29,29]& [-72,94] \\
    $f_{M_7}/\Lambda^4$ &[-13.3,12.9] & [-44,33]    &[-31,32]& [-79,107]\\
    $f_{T_0}/\Lambda^4$ &[-0.62,0.65] & [-1.35,1.60]  &[-3.7,3.8]  & [-8.0, 9.5] \\
    $f_{T_1}/\Lambda^4$ &[-0.28,0.31] & [-0.61,0.85]  &[-1.7,1.8]  & [-3.6, 5.0]\\
    $f_{T_2}/\Lambda^4$ &[-0.89,1.02] & [-2.1, 2.6]   &[-5.3,6.0]  & [-12, 15]\\
    \end{tabular}
    \end{center}
    \caption{Experimental limits (in TeV$^{-4}$) on dimension-8 operators 
      from the observation of $pp \rightarrow W^\pm W^\pm jjX$ by 
      CMS~\cite{Sirunyan:2017ret} (first column) and corresponding estimates 
      for the bounds on the Wilson coefficients $f_i/\Lambda^4$ in 
      the $T_u$ model (second column). Columns three and four give the 
      corresponding numbers for the VBFNLO normalization of operators. 
      See text for further details.}
    \label{table:newlimits}
    \end{table}

\begin{figure}[!bt] 
    \subfloat[Anomaly due to $\mathcal{O}_{S_1}$\label{subfig:mll_FS1}]{%
    \includegraphics[height= 6.55 cm, page = 1]{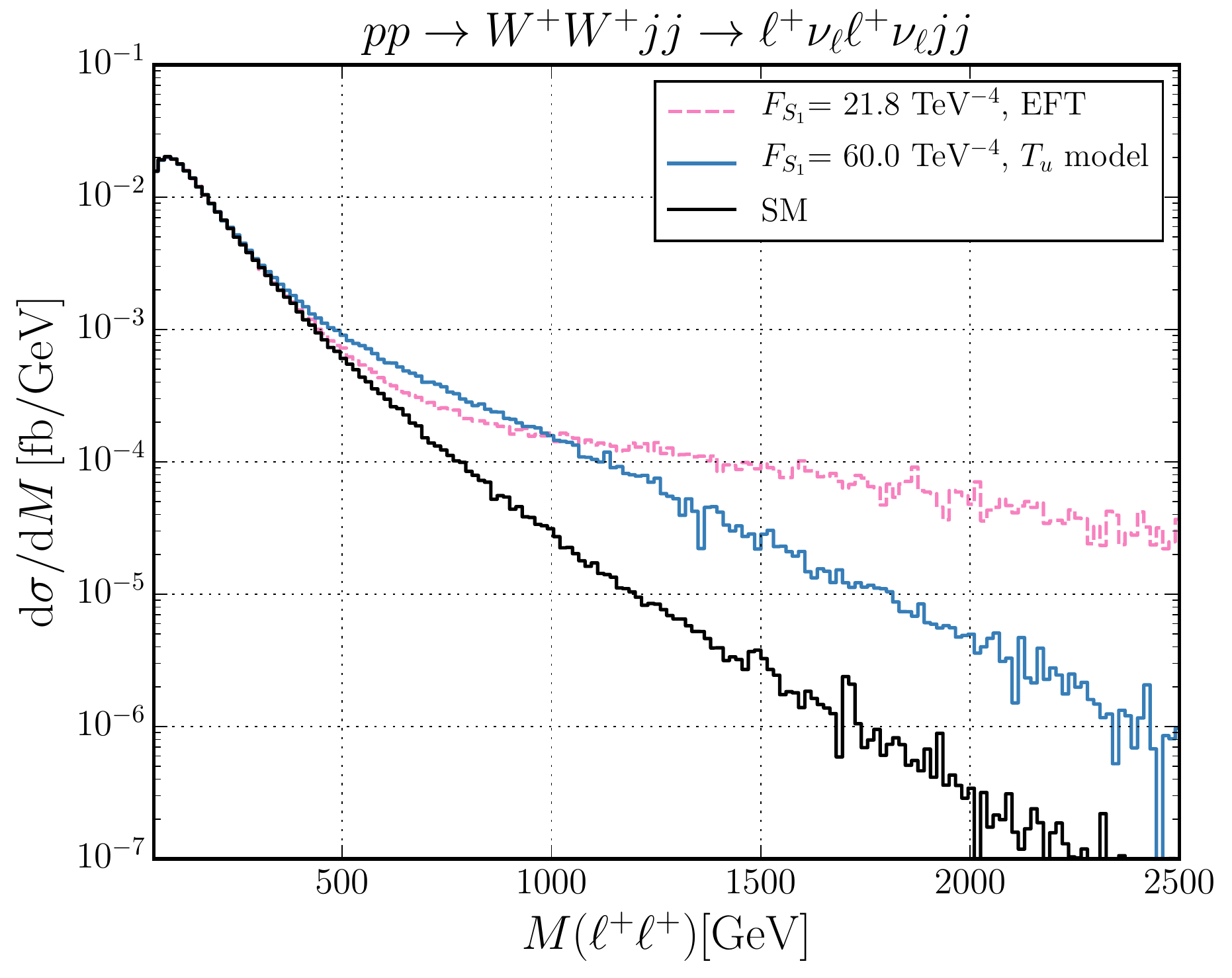}
    }
    \hfill 
    \subfloat[Anomaly due to $\mathcal{O}_{T_0}$\label{subfig:mll_FT0}]{%
    \includegraphics[height= 6.55 cm, page = 1]{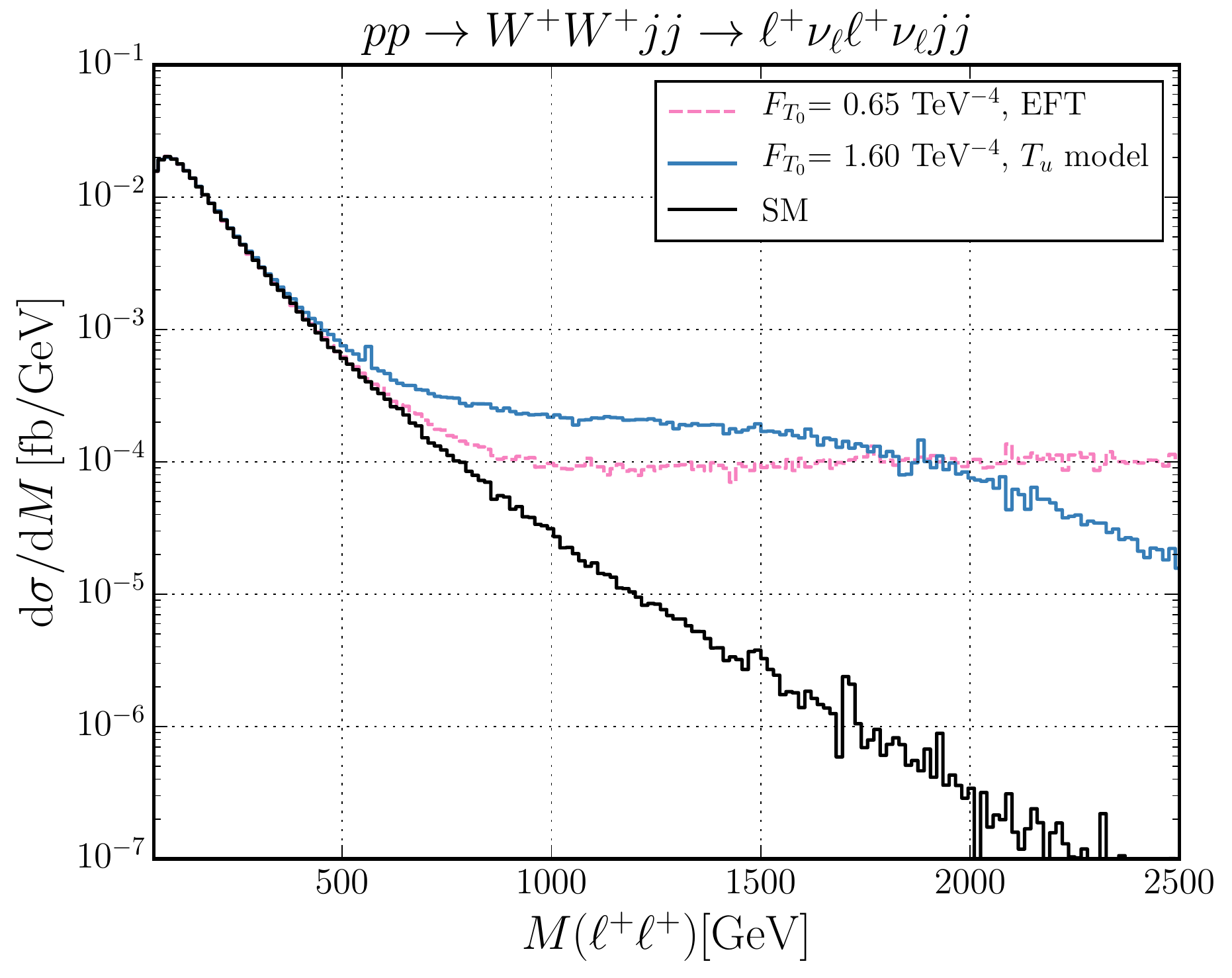}
    }
    \begin{center}
      \subfloat[Anomaly due to $\mathcal{O}_{M_0}$\label{subfig:mll_FM0}]{%
      \includegraphics[height= 6.55 cm, page = 1]{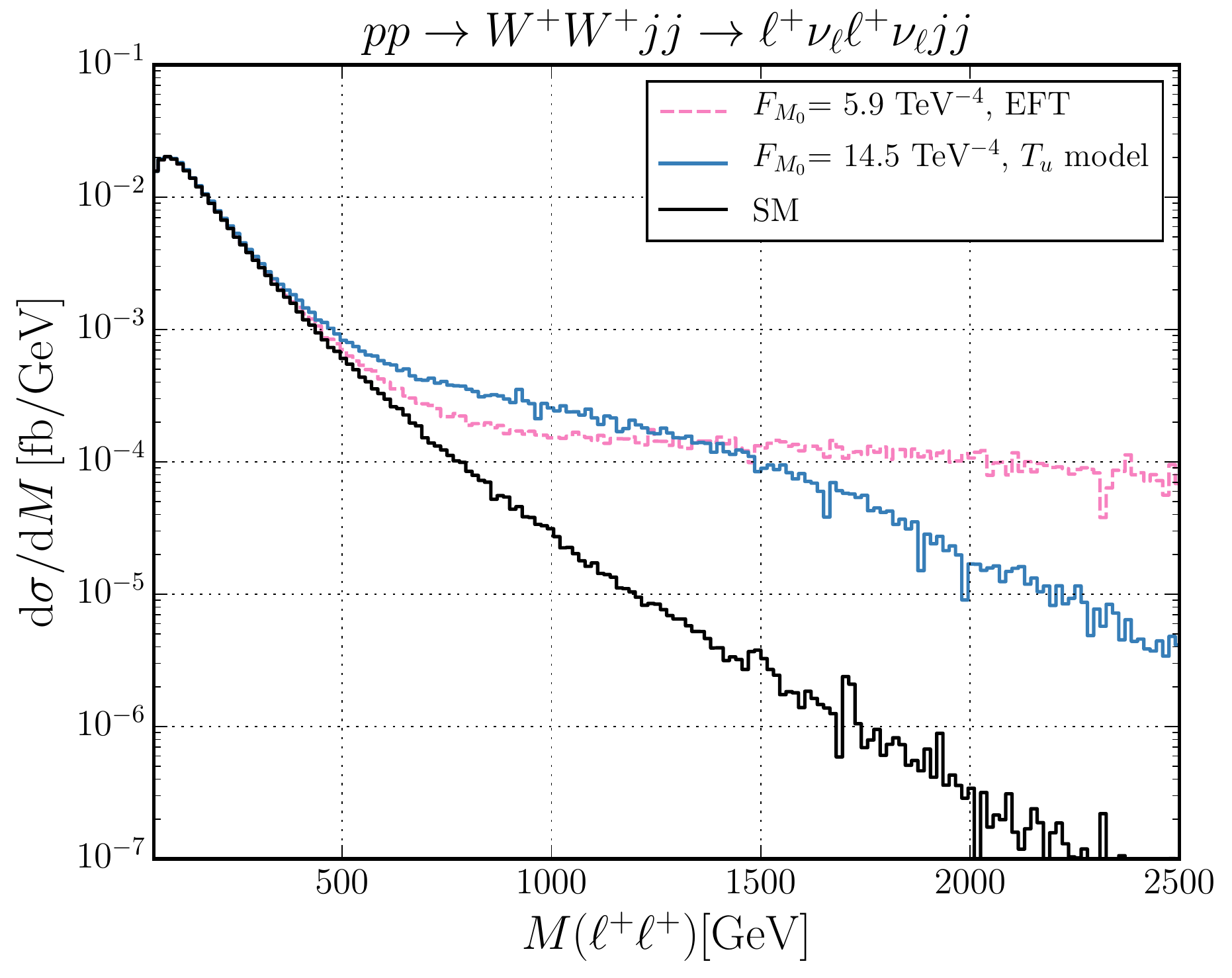}
      }
    \end{center} 
    \caption{ Differential cross section
    as a function of the invariant mass $m_{\ell \ell}$ of the charged leptons for
    $pp \rightarrow \ell^+ \nu_\ell \ell^+ \nu_\ell jjX$. 
    The solid black line shows the SM, while the dashed pink
    solid line shows the cross section for the anomalous coupling using
    the present CMS upper limit. The solid blue line shows the a
    unitarized curve with the same fiducial cross section as the EFT curve
    using the $T_u$ model. The fiducial region is defined in 
    Eq.~\eqref{eq:cuts}.}
     \label{fig:mll}
  \end{figure}

  The differential cross section as a function of the invariant mass 
  of the W-pair in Fig.~\ref{fig:compare_UnitModel} cannot be accessed
  experimentally, because the 4-momentum of the neutrinos is not measurable. A
  more readily accessible observable is the differential distribution in 
  the invariant mass of the two charged leptons, which is correlated to a 
  sufficient degree to the invariant mass of the two vector bosons. 
  Fig.~\ref{fig:mll} shows the corresponding 
  distribution for one non-zero coefficient of each class,
  namely the longitudinal $\mathcal{O}_{S_1}$ (Fig.~\ref{subfig:mll_FS1}), the
  transverse $\mathcal{O}_{T_0}$ (Fig.~\ref{subfig:mll_FT0}) and the mixed
  $\mathcal{O}_{M_0}$ (Fig.~\ref{subfig:mll_FM0}) operator. 
  For all three coefficients, the $T_u$ model is suppressed at high
  energy scales where the EFT description violates unitarity. However, the high
  energy tails differ by approximately one order of magnitude between the
  longitudinal operator $\mathcal{O}_{S_1}$ and the transverse operator
  $\mathcal{O}_{T_0}$. The differential cross section of the mixed operator in
  the the $T_u$  model lies between these. Below 500 GeV, the event production
  is mainly driven by SM contributions, as indicated by the fact that the SM 
  curve coincides with the 
  $T_u$ model curve for all three operators.

  In an attempt to distinguish the different operator types, we study the 
  transverse momentum of the leading tagging jet, $p_{T,\mathrm{max}}(j)$, and 
  the difference of the two lepton transverse momenta, 
  $\Delta p_{T,\ell \ell}=|{\bf p}_{T,\ell_1}-{\bf p}_{T,\ell_2}|$.  
  To optimize the ratio of BSM to SM events for the following study, 
  Fig.~\ref{fig:mll} suggests a cut, $m_{\ell^+\ell^+}>500$~GeV, on  the 
  charged lepton pair invariant mass. We show only the transverse and 
  the longitudinal operators in the following and omit the mixed 
  operators, which fall somewhat in between. 

\begin{figure}[bt] 
    \subfloat[Differential cross section\label{subfig:ptjmax}]{%
    \includegraphics[height= 6.55 cm, page = 1]{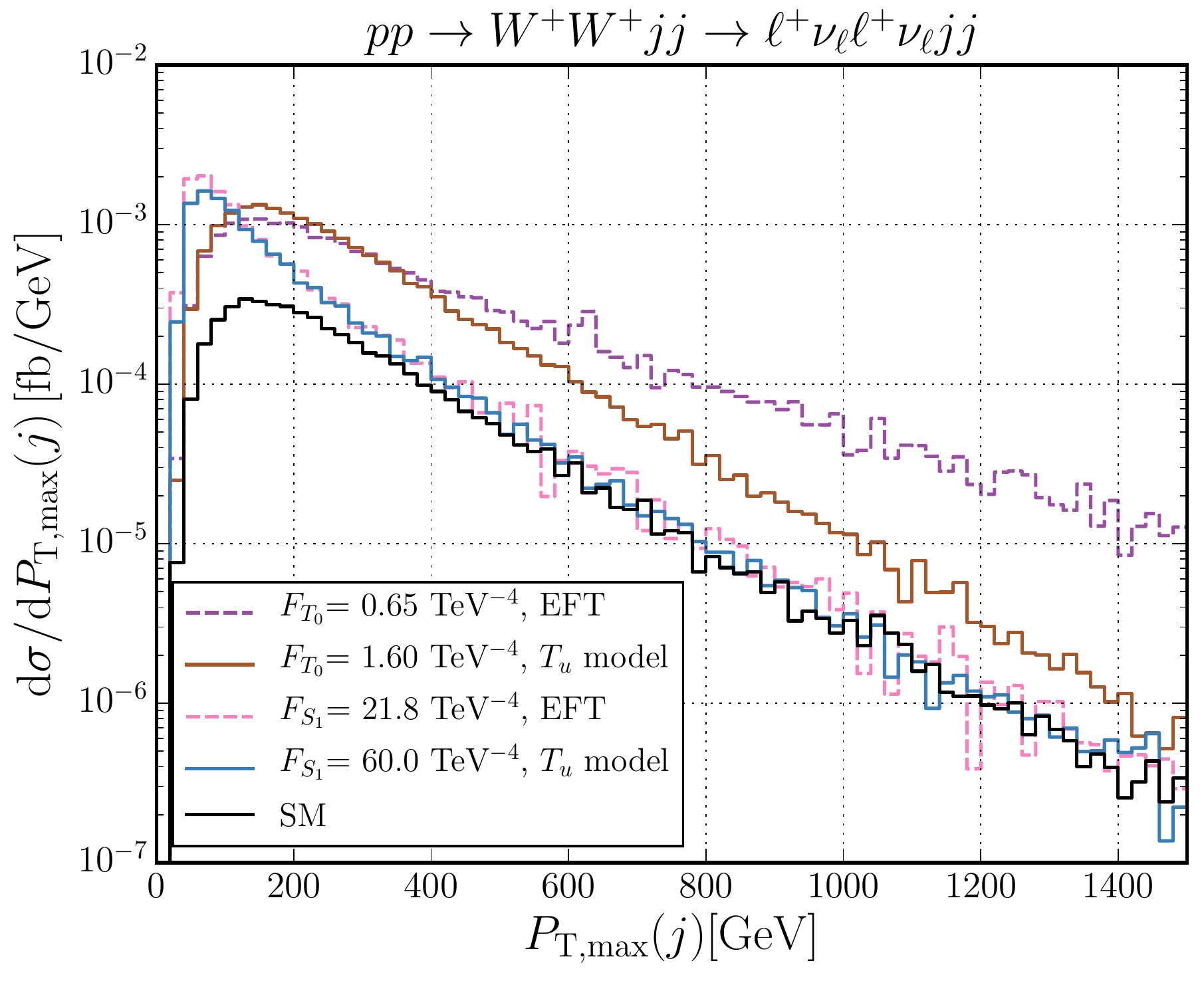}
    }
    \subfloat[Normalized distribution\label{subfig:ptjmax-rescaled}]{%
    \includegraphics[height= 6.55 cm, page = 1]{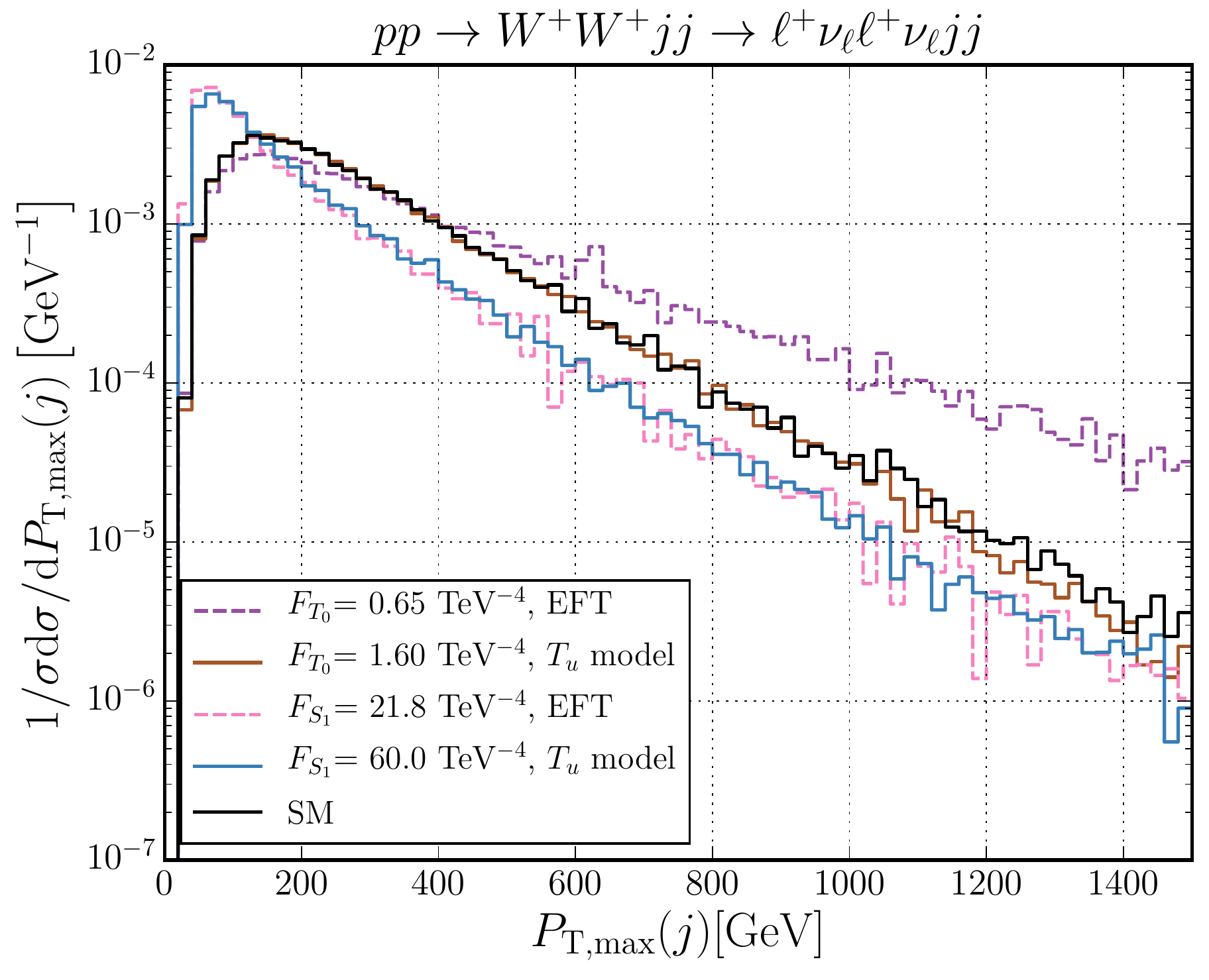}
    }
    \caption{Differential cross section
    as a function of $p_{T,\mathrm{max}}(j)$ for
    $pp \rightarrow \ell^+ \nu_\ell \ell^+ \nu_\ell jjX$. 
    The solid black line shows the SM, the dashed lines show the naive EFT 
    prediction within present CMS bounds, and the solid blue and brown lines 
    show the corresponding $T_u$ model for $\mathcal{O}_{S_1}$ and 
    $\mathcal{O}_{T_0}$. Beyond the cuts in Eq.~\eqref{eq:cuts} we impose 
    $m_{\ell \ell} > 500$ GeV.}
     \label{fig:ptjmax}
  \end{figure}

\begin{figure}[bt] 
    \subfloat[Differential cross section\label{subfig:ptll}]{%
    \includegraphics[height= 6.55 cm, page = 1]{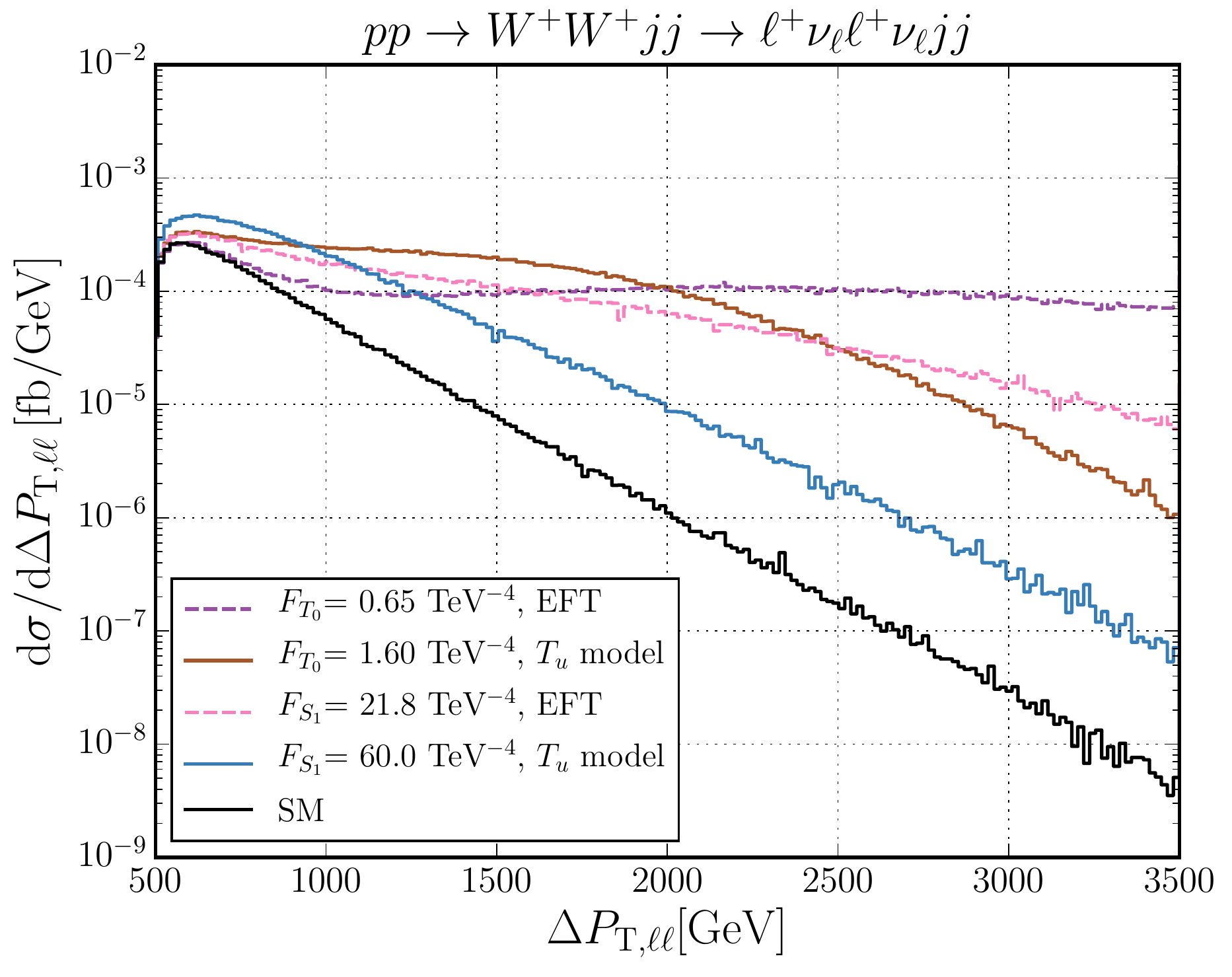}
    }
    \hfill 
    \subfloat[Normalized distribution\label{subfig:ptll-rescaled}]{%
    \includegraphics[height= 6.55 cm, page = 1]{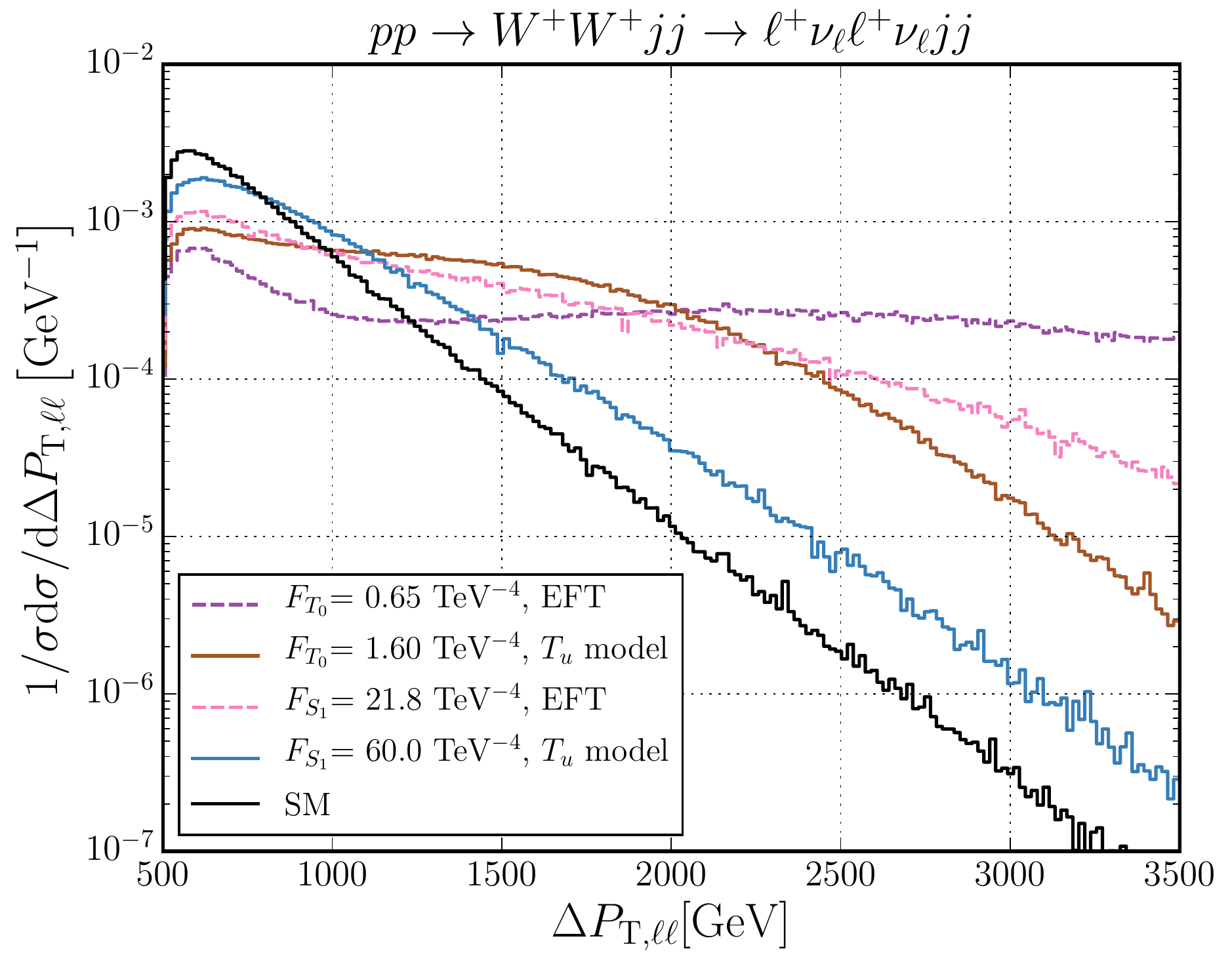}
    }
    \caption{Differential cross section
    as a function of $\Delta p_{T,\ell \ell}$ for
    $pp \rightarrow \ell^+ \nu_\ell \ell^+ \nu_\ell jjX$. 
    The solid black line shows the SM, the dashed lines show the naive EFT 
    prediction within current CMS bounds, and the solid blue and brown lines 
    show the corresponding $T_u$ model for $\mathcal{O}_{S_1}$ and 
    $\mathcal{O}_{T_0}$. Beyond the cuts in Eq.~\eqref{eq:cuts} we impose 
    $m_{\ell \ell} > 500$ GeV.}
     \label{fig:ptll}
  \end{figure}

  In Fig.~\ref{subfig:ptjmax} and Fig.~\ref{subfig:ptll},
  the differential cross sections as a function of $p_{T,\mathrm{max}}(j)$ and 
  $\Delta p_{T,\ell \ell}$, respectively, are plotted. 
  On the right-hand-side, in
  Fig.~\ref{subfig:ptjmax-rescaled} and Fig.~\ref{subfig:ptll-rescaled} the 
  same curves are shown as normalized distributions, which helps to better 
  expose differences in shape.  We
  compare the slope of the SM (solid black), the longitudinal operator
  $\mathcal{O}_{S_1}$ ($T_u$: solid blue, naive EFT: dashed pink) and the
  transverse operator $\mathcal{O}_{T_0}$ ($T_u$: solid brown, naive EFT: dashed
  purple).

Since incoming transversely polarized weak bosons lead to a harder jet $p_T$
distribution than longitudinally polarized bosons~\cite{Dawson:1984gx}, 
and since the transverse
operators enhance the transverse components,  we expect more events at larger
$p_{T,\mathrm{max}}(j)$  for the transverse operators as compared to the
longitudinal ones. This is clearly borne out in Fig.~\ref{fig:ptjmax}, which
shows a considerably harder $p_{T,\mathrm{max}}(j)$ spectrum for the
$\mathcal{O}_{T_0}$ operator than for $\mathcal{O}_{S_1}$. The cross section
enhancement for the longitudinal $\mathcal{O}_{S_1}$ operator occurs at small
$p_T$, which is typical for incident longitudinal bosons. At large
$p_{T,\mathrm{max}}(j)$, where incident transversely polarized $W$s dominate,
the SM and  $\mathcal{O}_{S_1}$ curves coincide, indicating that the underlying
anomalous quartic gauge coupling is mostly longitudinal.

Anomalous transverse operators produce cross section enhancements also at large
$p_{T,\mathrm{max}}(j)$. Here, an interesting difference can be observed between
the naive EFT model and our $T_u$ model: $T_u$ unitarization considerably
softens the $p_{T,\mathrm{max}}(j)$ spectrum (dashed purple to solid brown
curves). This effect is caused by the suppression of any large enhancement of
the $VV\to VV$ partial wave amplitudes, irrespective of its origin. For the
transverse operators one finds unphysically large enhancements also at high
virtualities of the incoming $W$s, while the $2\to 2$ center of mass energy,
$\sqrt{s}=m_{WW}$, remains small. Such an enhancement would not be corrected by
a unitarization attempt which relies only on suppression at large $s$, such as
the form-factor unitarization implemented previously in VBFNLO.\footnote{This
problem for a form-factor implementation can easily be cured by generalizing the
functional dependence of the form-factor, e.g. to 
$F(s,q_1^2,q_2^2) = (1+z (s^2+(q_1^2+q_2^2)^2)/\Lambda_{FF}^4)^n$, 
with $n\le -1$ and a phase factor $z=1$ or $z=i$.} Thus, one 
needs to be cautious when devising observables for transversely
polarized scattering based on a naive EFT approach: The large enhancement at
high $p_{T,\mathrm{max}}(j)$ for the purple $\mathcal{O}_{T_0}$ curve is an
artifact of the missing unitarization. The properly unitarized distribution has
a shape which is almost identical to the SM curve in
Fig.~\ref{subfig:ptjmax-rescaled}, which is also dominated by incoming
transversely polarized $W$s. Rather, the distinction between incoming
longitudinal and transverse weak bosons has to rely on the differences in the
$0<p_{T,\mathrm{max}}(j)<200$~GeV region,  where, fortunately, also the bulk of
the cross section is concentrated in all cases. 

A transversely polarized $W^+$ with helicity $\lambda=+1$ tends to emit the
charged anti-lepton in the forward direction relative to the $W$-momentum, which
leads to a high lepton $p_T$.  This is in contrast to a negatively polarized
$W^+$, which produces a relatively soft $\ell^+$, and the nearly equally shared
energy between $\ell^+$ and neutrino for the decay of an energetic,
longitudinally polarized $W$. Thus, $\Delta p_{T,\ell \ell}$ promises to
distinguish $(\lambda_3,\lambda_4)=(+,+)$ helicities (high $\Delta p_{T,\ell
\ell}$) from e.g. the $(0,0)$ or $(+,-)$ helicity combinations at lower average
$\Delta p_{T,\ell \ell}$. 

The corresponding differences are clearly exhibited in Fig.~\ref{fig:ptll}. Also
for the $\Delta p_{T,\ell \ell}$-distribu-tions, the slopes of the dimension-8
operator enhancements are noticeably influenced by the $T_u$ model. The
unphysical events at larger $\Delta p_{T,\ell \ell}$ are suppressed because
$\Delta p_{T,\ell \ell}$ and $\sqrt{s}=m_{WW}$ are highly correlated. The $T_u$
model prediction for the longitudinal operators and the SM have more events at
$\Delta p_{T,\ell \ell}$ below 1000 GeV and receive a large suppression for
larger $\Delta p_{T,\ell \ell}$. As expected, the transverse operator produces a
broader distribution, i.e. the enhancement due to $(\lambda_3,\lambda_4)=(+,+)$
polarization is clearly visible. For this observable, the unitarization model
even increases the discrimination power between different operators types.

\section{Discussion and Conclusions}
\label{sec:conclusions}

The parameterization of new physics effects in vector boson scattering via
anomalous quartic gauge couplings or an effective field theory, including
operators up to dimension-8, is a useful tool for analyzing VBS at the LHC.
However, because of the large energy reach of hadron colliders, which spans 
from low energies and momentum transfers where the pure EFT description is
valid, to regions of phase space where the polynomial growth of partial
wave amplitudes with energy exceeds unitarity limits, the naive EFT
description must be generalized to a model which respects unitarity bounds.
In this paper, we have developed the $T_u$ model, which is one such
generalization and which closely mirrors a K-matrix or linear T-matrix
unitarization of anomalous VBS amplitudes.

The $T_u$ model has been implemented as a purely numerical procedure in
the Monte Carlo program VBFNLO~\cite{Arnold:2008rz}, which allows to analyze VBS at NLO QCD 
precision, for arbitrary dimension-8 operators~\cite{Baglio:2014uba}. 
The $T_u$ model is constructed such that it 
reduces to the naive EFT approximation in all phase space regions where
this description is valid, and it smoothly interpolates to a unitarized
description for VBS at high virtualities. These high virtualities may either
correspond to high boson-pair invariant masses, $m_{V_3V_4}$, signified by
high energy and transverse momentum of the produced vector bosons
in $V_1V_2\to V_3V_4$, or to highly off-shell incoming $V_1$ or $V_2$,
i.e. large space-like $q_i^2$, which corresponds to
$pp\to V_3V_4jjX$ events with  tagging jets at very high transverse momentum.
Unphysical growth of VBS cross sections at high tagging jet $p_T$,
(see Fig.~\ref{fig:ptjmax}) which is present in a naive EFT implementation
even at small $m_{V_3V_4}$, also
needs to be suppressed, and the $T_u$ model does provide this regularization.

The purely numerical implementation grants great versatility and avoids
analytical approximations, like neglecting $m_W^2/s$ or $q_i^2/s$ suppressed
terms in a high energy approximation. It allows for arbitrary combinations
of dimension-8 operators to be present in the effective Lagrangian and thus
provides a general unitarized framework to analyze the effects of 
dimension-8 operators in VBS at the LHC. In addition, the numerical isolation
of off-shell $VV\to VV$ helicity amplitudes at intermediate steps of 
the calculation, allows, with little additional effort, to generate events for
selected center of mass helicities in the BSM $VV\to VV$ contribution, 
similar to a recent implementation in the PHANTOM 
Monte Carlo~\cite{Ballestrero:2017bxn,Ballestrero:2007xq}. 
So far, the implementation of the $T_u$ model has been tested and is
available for same-sign $W$-boson scattering, more precisely for
$pp\rightarrow W^\pm W^\pm jjX\rightarrow \ell^\pm \nu_\ell\ell^\pm \nu_\ell jjX$.
However, the generalization to single charged VBS ($WZ$-scattering) and neutral
channels will become available soon~\cite{Heiko2018}.

For same-sign $W$ scattering we have analyzed distributions which promise a
differentiation between individual tensor structures of the operators in
the EFT expansion, beyond
the only theoretically accessible di-boson invariant mass distribution 
in Fig.~\ref{fig:compare_UnitModel}  or the invariant mass distribution of the
two same-sign charged leptons in Fig.~\ref{fig:mll}. 
The transverse momentum distribution of the tagging
jets, e.g. $p_{T,\mathrm{max}}(j)$, which is shown in Fig.~\ref{fig:ptjmax}
is a good separator between longitudinal and transverse polarization of
the incident weak bosons. The charged lepton transverse momentum difference,
$\Delta p_{T,\ell\ell}$, which is shown in Fig.~\ref{fig:ptll}, can be used
to distinguish different combinations of $W$ polarizations in the final state.

For the same-sign $W$ case considered in this paper, we have shown in
Fig.~\ref{fig:compare_UnitModel} that the $T_u$ model closely agrees with the
T-matrix unitarization discussed by the WHIZARD group~\cite{Kilian:2014zja} for
longitudinal $W^+W^+$ scattering. However, the treatment of subleading,
$m_W^2/s$ or $q_i^2/s$ suppressed terms is different and means that the two
schemes provide different unitarization models. The numerical framework which is
now set up in the VBFNLO program allows for easy implementation of variants of
the $T_u$ model, such as taking into account more than just the largest
eigenvalue of the tree level scattering matrix for the denominator when going
from Eq.~(\ref{eq:T-matrix-offshell}) to Eq.~(\ref{eq:Tu-unitarisation}), or by
exploring other mappings of these real eigenvalues onto the Argand circle. We
leave such investigations to future work.  

\section*{Acknowledgments}
This work was supported in part by the BMBF Verbundforschung (HEP Theory).
The work of G.P. was supported by a fellowship of the
Karlsruhe School of Elementary Particle and Astroparticle Physics (KSETA).


\clearpage
\bibliographystyle{unsrt}
\bibliography{generalized-Tu-matrix}
 
\end{document}